\documentclass[11pt]{article}
\usepackage[dvips]{graphicx}
\usepackage{epsfig}
\usepackage{amssymb}
\usepackage{mathtools}
\usepackage{xcolor}

\def\PL#1#2#3{{Phys. Lett. }{\bf B#1 }(#2) #3}
\def\PLO#1#2#3{{Phys. Lett. }{\bf #1B }(#2) #3}

\def\PRV#1#2#3{{Phys. Rev. }{\bf #1 }(#2) #3}
\def\PRD#1#2#3{{Phys. Rev. }{\bf D#1 }(#2) #3}
\def\PRL#1#2#3{{Phys. Rev. Lett. }{\bf #1 }(#2) #3}
\def\NPB#1#2#3{{Nucl. Phys. }{\bf B#1 }(#2) #3}
\def\EPJ#1#2#3{{Eur. Phys. J. }{\bf C#1 }(#2) #3}
\def\CPC#1#2#3{{Comp. Phys. Comm. }{\bf #1 }(#2) #3}

\def\NIMA#1#2#3{{Nucl. Inst. Methods }{\bf A#1 }(#2) #3}

\def\be{\begin{equation}}
\def\ee{\end{equation}}
\def\vep{\varepsilon}
\def\ra{\rightarrow}
 
\newcommand{\PBO}{\mbox{\boldmath $\mathrm{p}$}}
\newcommand{\stat}{\mbox{$\mathrm{_{stat}}$}}
\newcommand{\syst}{\mbox{$\mathrm{_{syst}}$}}

\newcommand{\ext}{\mbox{$\mathrm{_{ext}}$}}

\newcommand{\expe}{\mbox{$\mathrm{_{exp}}$}}

\newcommand{\GEVcc}{\mbox{$\mathrm{GeV}/{{\it c}^2}$}}
\newcommand{\GEVc}{\mbox{$\mathrm{GeV}/{{\it c}}$}}
\newcommand{\GEV}{\mbox{$\mathrm{GeV}$}}
\newcommand{\MEVcc}{\mbox{$\mathrm{MeV}/{{\it c}^2}$}}
\newcommand{\MEVc}{\mbox{$\mathrm{MeV}/{\it c}$}}

\newcommand{\KPM}{\mbox{$K^\pm$}}
\newcommand{\PPI}{\mbox{$\pi^{+}$}}
\newcommand{\PMP}{\mbox{$\pi^{\pm}$}}
\newcommand{\MPI}{\mbox{$\pi^{-}$}}
\newcommand{\POP}{\mbox{$\pi^{+,0}$}}
\newcommand{\POM}{\mbox{$\pi^{-,0}$}}
\newcommand{\MEN}{\mbox{$M_{e\nu}$}}
\newcommand{\SE}{\mbox{$S_{e}$}}
\newcommand{\SL}{\mbox{$S_{\ell}$}}
\newcommand{\ZL}{\mbox{$z_{\ell}$}}
\newcommand{\MPP}{\mbox{$M_{\pi\pi}$}}
\newcommand{\SP}{\mbox{$S_{\pi}$}}
\newcommand{\PIo}{\mbox{$\pi^{0}$}}

\newcommand{\KPL}{\mbox{$K^{+}$}}
\newcommand{\KMI}{\mbox{$K^{-}$}}

\newcommand{\KTPO}{\mbox{$\mathrm{K}_{3\pi}^{00}$}}
\newcommand{\KLQ}{\mbox{$\mathrm{K}_{\ell 4}$}}
\newcommand{\KEQ}{\mbox{$\mathrm{K}_{\mathrm{e}4}$}}

\newcommand{\KEQP}{\mbox{$\mathrm{K}_{\mathrm{e}4}^{+}$}}
\newcommand{\KEQM}{\mbox{$\mathrm{K}_{\mathrm{e}4}^{-}$}}

\newcommand{\KEQO}{\mbox{$\mathrm{K}_{\mathrm{e}4}^{00}$}}
\newcommand{\KEQC}{\mbox{$\mathrm{K}_{\mathrm{e}4}^{+-}$}}

\newcommand{\KTE}{\mbox{$\mathrm{K}_{\mathrm{e}3}$}}

\newcommand{\ch}{\mbox{$\KPM \rightarrow \PPI \MPI e^{\pm} \nu$}}
\newcommand{\nt}{\mbox{$\KPM \ra \PIo \PIo e^{\pm} \nu$}}
\newcommand{\CUSP}{\mbox{$\KPM \rightarrow \PIo \PIo \PMP $}}

\newcommand{\KETC}{\mbox{$\KPM \ra \PIo e^{\pm} \nu$}}
\newcommand{\KDA}{\mbox{$\KPM \ra \PMP \PIo $}}

\newcommand{\PODK}{\mbox{$\pi^{0} \ra \gamma \gamma$}}
\newcommand{\PIEN}{\mbox{$\PMP \rightarrow e^{\pm} \nu $}}
\newcommand{\THP}{\mbox{$\theta_{\pi}$}}

\newcommand{\THL}{\mbox{$\theta_{\ell}$}}
\newcommand{\STP}{\mbox{$\sin\theta_{\pi}$}}
\newcommand{\STL}{\mbox{$\sin\theta_{\ell}$}}

\newcommand{\STTL}{\mbox{$\sin2\theta_{\ell}$}}
\newcommand{\CTP}{\mbox{$\cos\theta_{\pi}$}}
\newcommand{\CTL}{\mbox{$\cos\theta_{\ell}$}}
\newcommand{\CTE}{\mbox{$\cos\theta_{e}$}}
\newcommand{\CTTL}{\mbox{$\cos2\theta_{\ell}$}}
\newcommand{\CTTE}{\mbox{$\cos2\theta_{e}$}}
\newcommand{\CPH}{\mbox{$\cos\phi$}}
\newcommand{\SPH}{\mbox{$\sin\phi$}}

\newcommand{\SSTP}{\mbox{$\sin^2\theta_{\pi}$}}
\newcommand{\SSTE}{\mbox{$\sin^2\theta_{e}$}}
\newcommand{\SSTL}{\mbox{$\sin^2\theta_{\ell}$}}

\newcommand{\QRT}{\mbox{$\frac{1}{4}$}}
\newcommand{\HLF}{\mbox{$\frac{1}{2}$}}

\begin{document}
\centerline{\LARGE EUROPEAN ORGANIZATION FOR NUCLEAR RESEARCH}
\vspace{10mm} 
{\flushright{
CERN-PH-EP-2014-145 \\
July 18,2014 \\
\vspace{2mm}
}}
\vspace{15mm}

\begin{center}
{\bf {\Large \boldmath{Detailed study of the $\nt$ $(\KEQO)$ decay properties
\renewcommand{\thefootnote}{\fnsymbol{footnote}}\footnotemark[4]
}}}
\end{center}
\unboldmath
A sample of 65210 $\nt$ ($\KEQO$) decay candidates with 1\% background 
contamination has been collected in 2003--2004 by the NA48/2 collaboration
at the CERN SPS.
A study of the differential rate provides the first measurement of the hadronic
form factor variation in the plane $(M_{\pi\pi}^2 ,M_{e\nu}^2)$ and brings 
evidence for a cusp-like structure in the distribution of the squared 
$\PIo\PIo$ invariant mass around $4m_{\pi^+}^2$. 
Exploiting a model independent description of this form factor, the branching 
ratio, inclusive of radiative decays, is obtained using the $\CUSP$ decay mode
as normalization. It is measured to be BR$(\KEQO) = (2.552 
\pm 0.010\stat \pm 0.010\syst \pm 0.032\ext ) \times 10^{-5}$, which
improves the current world average precision 
by an order of magnitude while the
1.4\% relative precision is dominated by the external uncertainty from the 
normalization mode. A comparison with the properties of the corresponding 
mode involving a $\PPI \MPI$ pair ($\KEQC$) is also presented.

\noindent
\begin{center}
\it{Accepted for publication in JHEP}
\end{center}

\renewcommand{\thefootnote}{\fnsymbol{footnote}}
\footnotetext[4]{This study is dedicated to the memory of our colleague and friend Spasimir Balev $(1979-2013)$}
\renewcommand{\thefootnote}{\arabic{footnote}}

\newpage
\begin{center}
{\Large The NA48/2 Collaboration}\\
\vspace{2mm}
 J.R.~Batley,
 G.~Kalmus,
 C.~Lazzeroni$\,$\footnotemark[1]$^,$\footnotemark[2],
 D.J.~Munday,
 M.W.~Slater$\,$\footnotemark[1],
 S.A.~Wotton \\
{\em \small Cavendish Laboratory, University of Cambridge,
Cambridge, CB3 0HE, UK$\,$\footnotemark[3]} \\[0.2cm]
 R.~Arcidiacono$\,$\footnotemark[4],
 G.~Bocquet,
 N.~Cabibbo$\,$\renewcommand{\thefootnote}{\fnsymbol{footnote}}%
\footnotemark[2]\renewcommand{\thefootnote}{\arabic{footnote}},
 A.~Ceccucci,
 D.~Cundy$\,$\footnotemark[5],
 V.~Falaleev,
 M.~Fidecaro,
 L.~Gatignon,
 A.~Gonidec,
 W.~Kubischta,
 A.~Norton$\,$\footnotemark[6],
 A.~Maier,\\
 M.~Patel$\,$\footnotemark[7],
 A.~Peters\\
{\em \small CERN, CH-1211 Gen\`eve 23, Switzerland} \\[0.2cm]
 S.~Balev$\,$\renewcommand{\thefootnote}{\fnsymbol{footnote}}%
\footnotemark[2]\renewcommand{\thefootnote}{\arabic{footnote}},
 P.L.~Frabetti,
 E.~Gersabeck$\,$\footnotemark[8],
 E.~Goudzovski$\,$\footnotemark[1]$^,$\footnotemark[2],
 P.~Hristov$\,$\footnotemark[9],
 V.~Kekelidze,
 V.~Kozhuharov$\,$\footnotemark[10],
 L.~Litov$\,$\footnotemark[11],
 D.~Madigozhin,
 N.~Molokanova,
 I.~Polenkevich,\\
 Yu.~Potrebenikov,
 S.~Stoynev$\,$\footnotemark[12],
 A.~Zinchenko \\
{\em \small Joint Institute for Nuclear Research, 141980 Dubna (MO), Russia} \\[0.2cm]
 E.~Monnier$\,$\footnotemark[13],
 E.~Swallow,
 R.~Winston$\,$\footnotemark[14]\\
{\em \small The Enrico Fermi Institute, The University of Chicago,
Chicago, IL 60126, USA}\\[0.2cm]
 P.~Rubin$\,$\footnotemark[15],
 A.~Walker \\
{\em \small Department of Physics and Astronomy, University of Edinburgh, 
Edinburgh, EH9 3JZ, UK} \\[0.2cm]
 W.~Baldini,
 A.~Cotta Ramusino,
 P.~Dalpiaz,
 C.~Damiani,
 M.~Fiorini,
 A.~Gianoli,
 M.~Martini,
 F.~Petrucci,
 M.~Savri\'e,
 M.~Scarpa,
 H.~Wahl \\
{\em \small Dipartimento di Fisica e Scienze della Terra dell'Universit\`a e 
Sezione dell'INFN di Ferrara, I-44122 Ferrara, Italy} \\[0.2cm]
 A.~Bizzeti$\,$\footnotemark[16],
 M.~Lenti,
 M.~Veltri$\,$\footnotemark[17] \\
{\em \small Sezione dell'INFN di Firenze, I-50019 Sesto Fiorentino, Italy} \\[0.2cm]
 M.~Calvetti,
 E.~Celeghini,
 E.~Iacopini,
 G.~Ruggiero$\,$\footnotemark[9] \\
{\em \small Dipartimento di Fisica dell'Universit\`a e Sezione
dell'INFN di Firenze, I-50125 Sesto Fiorentino, Italy} \\[0.2cm]
 M.~Behler,
 K.~Eppard,
 M.~Gersabeck$\,$\footnotemark[18],
 K.~Kleinknecht,
 P.~Marouelli,
 L.~Masetti, \\
 U.~Moosbrugger,
 C.~Morales Morales$\,$\footnotemark[19],
 B.~Renk,
 M.~Wache,
 R.~Wanke,
 A.~Winhart$\,$\footnotemark[1] \\
{\em \small Institut f\"ur Physik, Universit\"at Mainz, D-55099
 Mainz, Germany$\,$\footnotemark[20]} \\[0.2cm]
 D.~Coward$\,$\footnotemark[21],
 A.~Dabrowski$\,$\footnotemark[9],
 T.~Fonseca Martin,
 M.~Shieh,
 M.~Szleper,\\
 M.~Velasco,
 M.D.~Wood$\,$\footnotemark[21] \\
{\em \small Department of Physics and Astronomy, Northwestern
University, Evanston, IL 60208, USA}\\[0.2cm]
 P.~Cenci,
 M.~Pepe,
 M.C.~Petrucci \\
{\em \small Sezione dell'INFN di Perugia, I-06100 Perugia, Italy} \\[0.2cm]
 G.~Anzivino,
 E.~Imbergamo,
 A.~Nappi$\,$\renewcommand{\thefootnote}{\fnsymbol{footnote}}%
\footnotemark[2]\renewcommand{\thefootnote}{\arabic{footnote}},
 M.~Piccini,
 M.~Raggi$\,$\footnotemark[10],
 M.~Valdata-Nappi \\
{\em \small Dipartimento di Fisica dell'Universit\`a e
Sezione dell'INFN di Perugia, I-06100 Perugia, Italy} \\[0.2cm]
 C.~Cerri,
 R.~Fantechi \\
{\em Sezione dell'INFN di Pisa, I-56100 Pisa, Italy} \\[0.2cm]
 G.~Collazuol$\,$\footnotemark[22],
 L.~DiLella,
 G.~Lamanna,
 I.~Mannelli,
 A.~Michetti \\
{\em Scuola Normale Superiore e Sezione dell'INFN di Pisa, I-56100
Pisa, Italy} \\[0.2cm]
 F.~Costantini,
 N.~Doble,
 L.~Fiorini$\,$\footnotemark[23],
 S.~Giudici,
 G.~Pierazzini$\,$\renewcommand{\thefootnote}{\fnsymbol{footnote}}%
\footnotemark[2]\renewcommand{\thefootnote}{\arabic{footnote}},
 M.~Sozzi,
 S.~Venditti$\,$\footnotemark[9] \\
{\em Dipartimento di Fisica dell'Universit\`a e Sezione dell'INFN di
Pisa, I-56100 Pisa, Italy} \\[0.2cm]
 B.~Bloch-Devaux$\,$\renewcommand{\thefootnote}{\fnsymbol{footnote}}%
\footnotemark[1]\renewcommand{\thefootnote}{\arabic{footnote}}$^,$\footnotemark[24],
 C.~Cheshkov$\,$\footnotemark[25],
 J.B.~Ch\`eze,
 M.~De Beer,
 J.~Derr\'e,
 G.~Marel,
 E.~Mazzucato,
 B.~Peyaud,
 B.~Vallage \\
{\em \small DSM/IRFU -- CEA Saclay, F-91191 Gif-sur-Yvette, France} \\[0.2cm]
\newpage
 M.~Holder,
 M.~Ziolkowski \\
{\em \small Fachbereich Physik, Universit\"at Siegen, D-57068
 Siegen, Germany$\,$\footnotemark[26]} \\[0.2cm]
 C.~Biino,
 N.~Cartiglia,
 F.~Marchetto \\
{\em \small Sezione dell'INFN di Torino, I-10125 Torino, Italy} \\[0.2cm]
 S.~Bifani$\,$\footnotemark[1],
 M.~Clemencic$\,$\footnotemark[9],
 S.~Goy Lopez$\,$\footnotemark[27] \\
{\em \small Dipartimento di Fisica dell'Universit\`a e
Sezione dell'INFN di Torino,\\ I-10125 Torino, Italy} \\[0.2cm]
 H.~Dibon,
 M.~Jeitler,
 M.~Markytan,
 I.~Mikulec,
 G.~Neuhofer,
 L.~Widhalm$\,$\renewcommand{\thefootnote}{\fnsymbol{footnote}}%
\footnotemark[2]\renewcommand{\thefootnote}{\arabic{footnote}} \\
{\em \small \"Osterreichische Akademie der Wissenschaften, Institut
f\"ur Hochenergiephysik,\\ A-10560 Wien, Austria$\,$\footnotemark[28]} \\[0.5cm]
\end{center}

\setcounter{footnote}{0}
\renewcommand{\thefootnote}{\fnsymbol{footnote}}
\footnotetext[1]{Corresponding author, email: brigitte.bloch-devaux@cern.ch}
\footnotetext[2]{Deceased}
\renewcommand{\thefootnote}{\arabic{footnote}}
\footnotetext[1]{Now at: School of Physics and Astronomy, University of Birmingham,  Birmingham, B15 2TT, UK}
\footnotetext[2]{
UF100308, UF0758946}
\footnotetext[3]{Funded by the UK Particle Physics and Astronomy Research Council, grant PPA/G/O/1999/00559}
\footnotetext[4]{Now at: Universit\`a degli Studi del Piemonte Orientale e Sezione 
dell'INFN di Torino, I-10125 Torino, Italy}
\footnotetext[5]{Now at: Istituto di Cosmogeofisica del CNR di Torino,
I-10133 Torino, Italy}
\footnotetext[6]{Now at: Dipartimento di Fisica e Scienze della Terra dell'Universit\`a e Sezione
dell'INFN di Ferrara, I-44122 Ferrara, Italy}
\footnotetext[7]{Now at: Department of Physics, Imperial College, London,
SW7 2BW, UK}
\footnotetext[8]{Now at: Physikalisches Institut, Ruprecht-Karls-Universit\"at Heidelberg, D-69120 Heidelberg, Germany}
\footnotetext[9]{Now at: CERN, CH-1211 Gen\`eve 23, Switzerland}
\footnotetext[10]{Now at: Laboratori Nazionali di Frascati, I-00044 Frascati, Italy}
\footnotetext[11]{Now at: Faculty of Physics, University of Sofia ``St. Kl.
Ohridski'', 1164 Sofia, Bulgaria, 
funded by the Bulgarian National Science Fund under contract DID02-22}
\footnotetext[12]{Now at: Northwestern University,
Evanston, IL 60208, USA}
\footnotetext[13]{Now at: Centre de Physique des Particules de Marseille,
IN2P3-CNRS, Universit\'e de la M\'editerran\'ee, F-13288 Marseille,
France}
\footnotetext[14]{Now at: School of Natural Sciences, University of California, Merced, CA 95343, USA}
\footnotetext[15]{Now at: School of Physics, Astronomy and Computational Sciences, George Mason
University, Fairfax, VA 22030, USA}
\footnotetext[16]{Also at Dipartimento di Fisica, Universit\`a di Modena e
Reggio Emilia, I-41125 Modena, Italy}
\footnotetext[17]{Also at Istituto di Fisica, Universit\`a di Urbino,
I-61029 Urbino, Italy}
\footnotetext[18]{Now at: School of Physics and Astronomy, The University of Manchester, Manchester, M13 9PL, UK}
\footnotetext[19]{Now at: Helmholtz-Institut Mainz, Universit\"at Mainz,
D-55099 Mainz,
Germany}
\footnotetext[20]{Funded by the German Federal Minister for
Education and research under contract 05HK1UM1/1}
\footnotetext[21]{Now at: SLAC, Stanford University, Menlo Park, CA 94025,
USA}
\footnotetext[22]{Now at: Dipartimento di Fisica dell'Universit\`a e Sezione dell'INFN di Padova, I-35131 Padova, Italy}
\footnotetext[23]{Now at: Instituto de F\'{\i}sica Corpuscular IFIC,
Universitat de Val\`{e}ncia, E-46071 Val\`{e}ncia, Spain}
\footnotetext[24]{Now at: Dipartimento di Fisica dell'Universit\`a di Torino, 
I-10125 Torino, Italy}
\footnotetext[25]{Now at: Institut de Physique Nucl\'eaire de Lyon,
IN2P3-CNRS, Universit\'e Lyon I, F-69622 Villeurbanne, France}
\footnotetext[26]{Funded by the German Federal Minister for Research
and Technology (BMBF) under contract 056SI74}
\footnotetext[27]{Now at: Centro de Investigaciones Energeticas
Medioambientales y Tecnologicas, E-28040 Madrid, Spain}
\footnotetext[28]{Funded by the Austrian Ministry for Traffic and
Research under the contract GZ 616.360/2-IV GZ 616.363/2-VIII, and
by the Fonds f\"ur Wissenschaft und Forschung FWF Nr.~P08929-PHY}


\clearpage


\section{Introduction}
Kaon decays involve weak, electromagnetic and strong interactions in an 
intricate mixture but also result in experimentally simple low multiplicity 
final states. Because of the small kaon mass value, 
these decays have been identified as a perfect 
laboratory to study hadronic low energy processes away from the multiple-pion
resonance region.
Semileptonic four-body $\KPM$ decays ($\KPM \rightarrow \pi \pi l^{\pm} \nu$
 denoted $\KLQ$) are of particular interest
because of the small number of hadrons in the final state and the 
well-understood Standard Model electroweak amplitude responsible for the 
leptonic part.
In the non-perturbative QCD regime at such low energies (below 1 $\GEV$), 
the development over more than 30 years of chiral perturbation theory (ChPT) 
\cite{chpt1} and more recently of lattice QCD \cite{lattice} has reached in 
some domains a precision level competitive with the most accurate experimental 
results.

The global analysis of $\pi\pi$ and $\pi K$ scattering and $\KLQ$ decay data 
allows for the determination of the Low Energy Constants (LEC) of ChPT at Leading 
and Next to Leading Orders
\cite{BCG,BJ} and subsequent predictions of form factors and decay rates. 
The possibility to study high statistics samples collected concurrently by 
NA48/2 in several modes brings improved precision inputs and therefore allows 
stringent tests of ChPT predictions.

A total of 37 $\nt$ ($\KEQO$) decays were observed several decades ago  
by two experiments in heavy liquid bubble chamber exposures to $\KPL$ beams 
\cite{cline,barmin}, and a counter experiment using a $\KMI$ beam 
\cite{istra}. At that time, in the framework of current algebra and under the 
assumption of a unique and constant form factor $F$ \cite{berends}, the 
$\KEQO$ decay rate and form factor values were related by 
$\Gamma = (0.75 \pm 0.05) ~| V_{us} \cdot F|^{2} 10^{3} {\rm ~s}^{-1}$.
In 2004, the E470 experiment at KEK \cite{K470} reported an observation of 214
candidates in a study of stopped kaon decays in an active target. Due to a 
very low geometrical acceptance and large systematics, the partial rate 
measurement did not reflect the gain in statistics and was not included in the 
most recent world averages of the Particle Data Group \cite{pdg}, BR = 
$(2.2 \pm 0.4) \times 10^{-5}$, unchanged since the 1990's, corresponding to 
the model dependent form factor value $| V_{us} \cdot F| = 1.54 \pm 0.15$.

The detailed analysis of more than one million events in the ``charged pion'' 
$\KEQ$ decay mode ($\ch$ denoted $\KEQC$) \cite{ke410,ke412} is now 
complemented by the analysis of a large sample in the ``neutral pion'' $\KEQO$ 
decay mode.
This sample (65210 $\KEQO$ decays with 1\% background), though not as large as
the $\KEQC$ sample, is larger than the 
total world sample by several orders of magnitude. A control of the systematic 
uncertainties competitive with the statistical precision allows both form 
factor and rate, using the $\CUSP$ ($\KTPO$) decay mode as normalization,
to be measured with improved precision. These  model independent measurements 
and a discussion of their possible interpretation are reported here.

\section{The NA48/2 experiment beam and detector}
\label{sec:beamdet}
The NA48/2 experiment was specifically designed for charge asymmetry
measurements in $\KPM$ decays to three pions~\cite{agcomb} taking advantage 
of simultaneous $\KPL$ and $\KMI$ beams produced by $400~\GEVc$ primary CERN 
SPS protons impinging on a 40~cm long beryllium target. 
Oppositely charged particles ($p,\pi,K$), with a central momentum of
$60~\GEVc$ and a momentum band of $\pm 3.8\%$ (rms), are selected by 
two systems of dipole magnets with zero total deflection (each of them forming 
an `achromat'), focusing quadrupoles, muon sweepers, and collimators.

At the entrance of the decay volume enclosed  in a 114 m long 
vacuum tank, the beams contain $\sim2.3\times10^6~\KPL$ and
$\sim1.3\times10^6~\KMI$ per pulse of about 4.5 s duration.
Both beams follow the same path in the decay 
volume: their axes coincide within 1~mm, while the transverse size of the beams
is about 1~cm. The fraction of beam kaons decaying in the vacuum tank at 
nominal momentum is about $22\%$.

The decay volume is followed by a magnetic spectrometer 
housed in a tank filled with helium at nearly atmospheric pressure, separated 
from the vacuum tank by a thin 
($\sim 0.4\%X_0$) $\rm{Kevlar}\textsuperscript{\textregistered}$ composite 
window. An aluminum beam pipe of 8~cm outer radius and 1.1 mm thickness,
traversing the centre of the spectrometer (and all the following
detector elements), allows the undecayed beam particles and the muon
halo from decays of beam pions to continue their path in vacuum. The
spectrometer consists of four octagonal drift chambers (DCH), each composed of 
four staggered double planes of sense wires, located upstream (DCH1--2) and 
downstream (DCH3--4) of a large aperture dipole magnet. 
The magnet provides a transverse momentum kick $\Delta p=120~\MEVc$ to 
charged particles in the horizontal plane. The spatial resolution of each DCH 
is $\sigma_x=\sigma_y=90~\mu$m and the momentum resolution achieved in the 
spectrometer is $\sigma_p/p = (1.02 \oplus 0.044\cdot p)\%$ ($p$ in $\GEVc$).

The spectrometer is followed by a hodoscope (HOD) consisting of two planes of 
plastic scintillator segmented into vertical and horizontal strip-shaped 
counters (128 in total). The HOD surface is logically subdivided into 
$2 \times 4$ exclusive square regions.
The time coincidence of signals in the two HOD planes in corresponding regions
define quadrants
whose fast signals are used to trigger the detector readout on charged track 
topologies. The achieved time resolution is $\sim 150$ ps.

A liquid krypton electromagnetic calorimeter (LKr), located behind the HOD, is 
used to reconstruct $\PODK$ decays and for particle identification in the 
present analysis. It is an almost homogeneous ionization chamber with an active
volume of 7 m$^3$ of liquid krypton, segmented transversally into 13248 
projective cells, approximately 2$\times$2 cm$^2$ each, $27X_0$ deep and 
without longitudinal segmentation. 
The energies of electrons and photons are measured with a resolution 
$\sigma_E /E = (3.2/\sqrt{E} \oplus 9.0/E \oplus 0.42)\%$ ($E$ in $\GEV$) and
the transverse position of isolated showers is measured with a
spatial resolution $\sigma_x = \sigma_y = (0.42/\sqrt{E} \oplus 0.06)$~cm. 

A hadron calorimeter and a muon veto counter are located further 
downstream. Neither of them is used in the present analysis.
A more detailed description of the NA48 detector and its performances can be 
found in Ref.~\cite{na48det}. 

The experiment collected a total of $1.8 \times 10^{10}$ triggers in two years 
of data-taking using a dedicated two-level trigger logic to 
select and flag events. In this analysis, only a specific trigger branch is
considered: at the first level, the trigger requires a signal in at least one
HOD quadrant (Q1) in coincidence with the presence of
energy depositions in LKr consistent with at least two photons (NUT).
At the second level (MBX), an on-line processor receiving the DCH 
information reconstructs the momentum of charged particles and calculates the
missing mass under the assumption that the particles are
$\PMP$ originating
from the decay of a $60~\GEVc ~\KPM$ traveling along the nominal beam axis.
The requirement that the missing mass $M_{miss}$ is larger than the 
$\PIo$ mass rejects most $\KDA$ decays (the lower trigger 
cutoff was $194~\MEVcc$ in 2003 and  $181 ~\MEVcc$ in 2004).

In $\KTPO$ decays $M_{miss}$ corresponds to the 
$\PIo \PIo$ system, with the minimum value of $2 m_{\pi^0}$ and 
satisfies the trigger requirement, while in $\KEQO$ decays
$M_{miss}$ can extend to much lower values (even to negative $M_{miss}^2$ 
values) because of the low electron mass. For this reason, $\sim 55 \%$ of 
$\KEQO$ decays are excluded at trigger level. In particular, the low momentum 
electron spectrum below 6 $\GEVc$ is totally excluded by this trigger 
condition.

\section{Measurement principle}
The $\KEQO$ rate is measured relative to the abundant $\KTPO$ normalization 
channel. As the topologies of the two modes are similar in terms of number of 
detected charged ($e^\pm$ or  $\pi^\pm $) and neutral (two $\PIo$-mesons, 
each decaying to $\gamma\gamma$) particles, the two samples are collected 
concurrently using the same trigger logic and a common selection is employed 
as far as possible. 
This leads to partial cancellation of the systematic
effects induced by imperfect kaon beam description, local detector
inefficiencies, and trigger inefficiency, and avoids relying on the absolute 
kaon flux measurement.
The ratio of the partial rates -- or branching ratios (BR) -- is obtained as:
\be
\Gamma(\KEQO)/\Gamma(\KTPO)  = {\rm BR}(\KEQO) / {\rm BR}(\KTPO) = \frac{N_{s} - N_{b}(s)}{N_{n} - N_{b}(n)} \cdot \frac{A_{n} ~\vep_{n}}{A_{s}~\vep_{s} }
\label{eq:br}
\ee
where $N_{s}, N_{n}$ are the numbers of signal and normalization candidates;  
$N_{b}(s), N_{b}(n)$ are the numbers of background events in the signal and 
normalization samples; $A_{s}$ and  $\vep_{s}$ are the geometrical
acceptance and trigger efficiency for the signal sample; $A_{n}$ and $\vep_{n}$
are those of the normalization sample.
The normalization branching ratio value BR$($\KTPO$) = (1.761 \pm 0.022)\%$ is 
the world average as computed in Ref.~\cite{pdg}.

As the geometrical acceptances are not uniform over the kinematic space, 
their overall values depend on the knowledge of the dynamics which 
characterizes each decay. This motivates a detailed study of the $\KEQO$ form 
factor in the kinematic space, never performed so far due to 
the very small size of the available samples.
Such a measurement will allow a model independent determination of the
branching ratio.

Due to different data taking conditions, acceptances (Section \ref{sec:accep})
and trigger efficiencies (Section \ref{sec:trig}) are not uniform over the
whole data sample. For this reason, ten independent subsamples
recorded with stable conditions are analyzed separately and statistically
combined to obtain the BR value.

\section{Event selection and reconstruction}
\label{sec:selevt}
The event selection and reconstruction follow as much as possible the same path
for both $\KTPO$ and $\KEQO$ samples and the separation between signal and
normalization occurs only at a later stage.

{\bf Common selection}.
Events are considered if at least four clusters are reconstructed in the LKr, 
each of them consistent
with the electromagnetic shower produced by a photon of energy above $3~\GEV$.
The distance between any two photons in the LKr is required to be larger than 
10 cm to minimize the effect of shower overlap.
Fiducial cuts on the distance of each photon 
from the LKr borders and central hole are applied to ensure full 
containment of the electromagnetic showers. 
In addition, because of the presence of $\sim100$ LKr cells affected by readout
problems ({\it inactive} cells), the minimum distance between the photon and 
the nearest LKr {\it inactive} cell is required to be at least 2 cm. 

Each possible pair of photons is assumed to originate 
from a $\PIo \ra \gamma \gamma$ decay and the distance $D_{ij}$ between 
the $\PIo$ decay vertex and the LKr front face ($Z_{LKr}$) is 
computed\footnote{The small angle approximation is satisfied by the detector 
geometry.}:
\vspace{-7mm}
\begin{center}
   $D_{ij} = \sqrt{E_{i}E_{j}} ~R_{ij} / m_{0}$,
\end{center}
\vspace{-8mm}
where $E_{i},E_{j}$ are the energies of the $i$-th and $j$-th photon, 
respectively, $R_{ij}$ is the distance between their impact points on the LKr, 
and $m_0$ is the $\PIo$ mass. Among all possible $\PIo$ pairs, only those with 
$D_{ij}$ and $D_{kl}$ values differing by less than 500 cm are retained further
(the rms of this distribution is $\sim150$ cm), and
the distance of the $\KPM$ decay vertex from the LKr is taken as the 
arithmetic average of the two $D_{ij}$ and $D_{kl}$ values. The longitudinal 
position along the beam axis of the {\it neutral vertex} is defined as 
$Z_n = Z_{LKr} - (D_{ij} + D_{kl})/2$. 
A further constraint is applied on the time difference between the 
earliest or latest 
cluster time and the four photon average time at $\pm 2.5$ ns,
taking advantage of the good time resolution of the calorimeter for photon 
clusters ($\sigma_t = 2.5~{\rm  ns} / \sqrt{E}$  (E in $\GEV$) \cite{na48det}).

A photon emitted at small angle to the beam 
axis may cross the aluminum vacuum tube in the spectrometer or the DCH1 
central flange before reaching the LKr. In such a case the photon 
energy may be mis-measured.
Therefore, the distance of each candidate photon to the 
nominal beam axis at the DCH1 plane is required to be larger than 11 cm 
(largest radial extension of the flange),
assuming an origin on axis at $Z_n + 400$ cm (this takes into account the
resolution of the $Z_n$ measurement of $\sim$ 80 cm). 

Events with at least one charged particle track having a momentum above 
$5~\GEVc$ and satisfying good quality reconstruction criteria are further 
considered.
The track coordinates should be within the fiducial acceptance of DCH1 
(distance from the beam axis $R >12$ cm) and HOD ($R >15$ cm) and outside the 
inefficient HOD areas. To ensure a uniform Q1 trigger efficiency at the first 
level trigger, two half slabs of the hodoscope affected by an intermittent 
hardware failure (one in part of the 2003 data, a different one in part of the 
2004 data) have been temporarily removed from the geometrical acceptance of the
event selection.
This track should also satisfy the 
requirement $M_{miss} > 206~\MEVcc$, more restrictive than the on-line second
level trigger cut and ensuring a 
high MBX trigger efficiency. 
The track impact at the LKr front face should be within the fiducial acceptance
and away from the closest {\it inactive} cell by more than 2 cm.
For each track candidate, the {\it charged vertex} longitudinal position $Z_c$ 
is defined at the closest distance of approach 
to the kaon beam axis, which in turn has to be smaller than 5 cm.
In addition, the distance between each photon candidate and the impact point of
the track on the LKr front face must exceed 15 cm.  
The track and four photon time difference must be consistent with the 
same decay within $\pm 15$ ns if using the DCH time or within $\pm 2.5$ ns if 
using the more precise HOD time (about $0.4\%$ of these tracks cannot be 
associated to a reliable HOD time).  

At the following step of the selection, the consistency of the surviving
events with the decay hypothesis of a kaon into one charged track and two 
$\PIo$-mesons is checked. The track candidate is kept if the $Z_c$ and $Z_n$ 
values are compatible within $\pm~800$ cm.
The rms $(\sigma_{D},\sigma_{nc})$ of the distributions $(D_{ij} - D_{kl})$ and
$(Z_n - Z_c)$ have been studied as a function of the neutral vertex position 
for selected candidates. They vary slowly with the $Z_n$ position and are 
parameterized by degree-2 polynomial functions.
If several tracks and $\PIo$ pairs satisfy the vertex criteria, the choice is 
made on the basis of the best geometrical vertex matching,
keeping the combination 
with the smallest value of the estimator 
$((D_{ij} - D_{kl})/ \sigma_{D})^2 + ((Z_n - Z_c)/ \sigma_{nc})^2$.
Up to this stage, both signal and normalization events follow the same 
selection and only one track-$\PIo$ pair combination per event is kept 
(96$\%$ of the candidates have a single combination). 

The reconstructed neutral vertex position
is further required to be located within a 106 m long fiducial volume contained
in the vacuum tank and starting 4 m downstream of the final beam collimator 
(to exclude $\PIo$-mesons produced from beam particles interacting in the 
collimator material).

{\bf Event reconstruction}. Each candidate is reconstructed in the plane
 $(M_{3\pi},p_t)$ where $M_{3\pi}$ is the  invariant mass of the three pion 
system (in the $\PIo \PIo \PMP$ hypothesis, giving a $\pi^+$ mass to the 
charged track) and $p_t$ is its transverse 
momentum relative to the mean nominal beam axis. 

The parent kaon momentum is reconstructed under two assumptions:
either as the total momentum sum of the charged track and the two $\PIo$-mesons
or imposing energy-momentum 
conservation in a four-body decay $\KEQ$ hypothesis (an electron mass is given 
to the charged track) and fixing the kaon mass and the beam direction to their 
nominal values. In the latter case, a quadratic equation in the kaon momentum 
$p_{K}$ is obtained and the solution closest to the nominal value is kept.

{\bf Particle identification}. Criteria are based on the geometric
association of an in-time LKr energy deposition cluster to a track extrapolated
to the calorimeter front face (denoted ``associated cluster'' below). The ratio
of energy deposition in the LKr calorimeter to momentum measured by the
spectrometer ($E/p$) is used for pion/electron separation. 
A track is identified as an {\it electron} (e$^{\pm}$) if its momentum is 
greater than $5~\GEVc$ and it has an associated cluster with $E/p$ between 0.9 
and 1.1.
A track is identified as a {\it pion} ($\pi^{\pm}$) if its momentum is above
$5~\GEVc$ (there is no requirement of an associated cluster). 

\begin{figure}[htb]
\begin{center}
\begin{picture}(160,70)
\put(-140,-130){\epsfxsize80mm\epsfbox{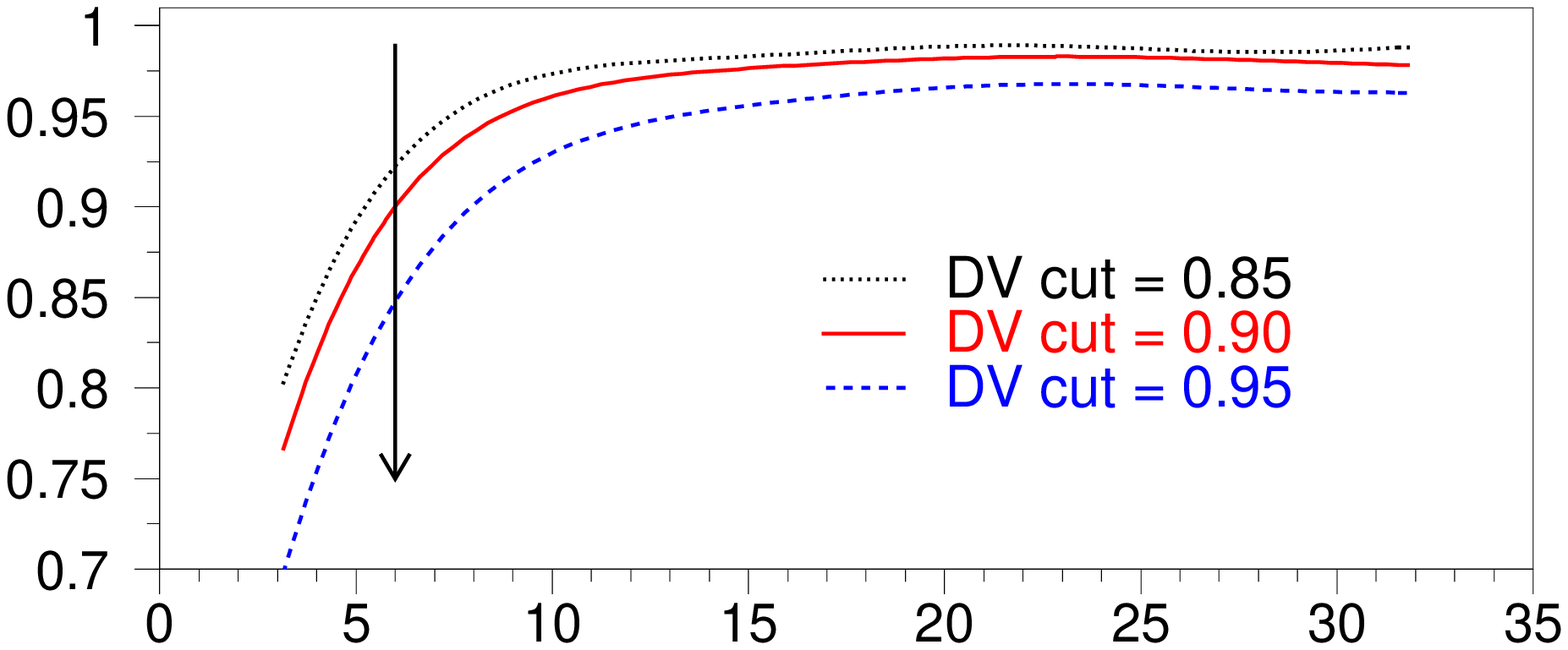}}
\put(-110,46){\bf \large (a)}
\put(20,-39){$p~(\GEVc)$}
\put(90,-130){\epsfxsize80mm\epsfbox{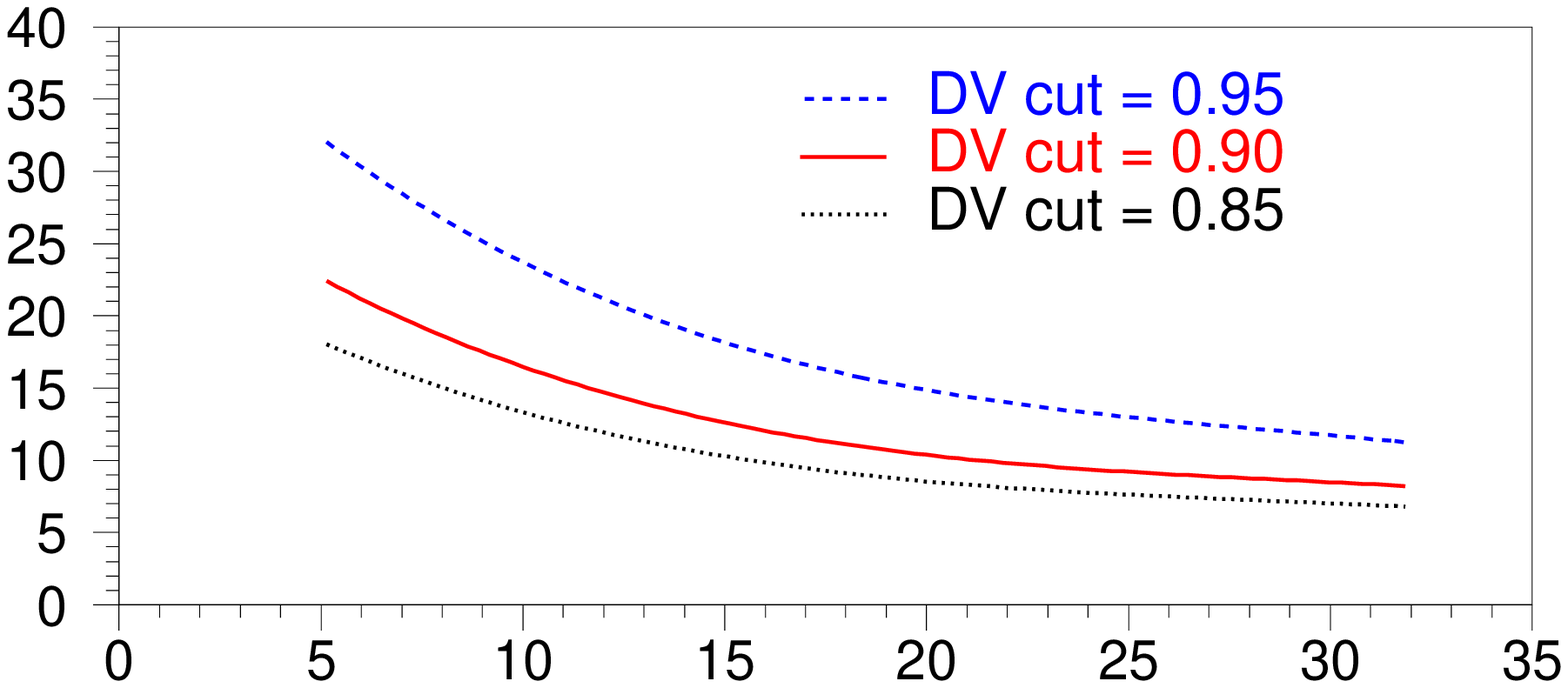}}
\put(120,46){\bf \large (b)}
\put(250,-39){$p~(\GEVc)$}
\end{picture}
\end{center}
\caption{\label{fig:lda} 
{\bf (a)} Efficiency of the discriminating variable as a function of momentum 
for electron tracks and three DV cut values. The arrow shows the effective 
minimum electron momentum in the signal selection.
{\bf (b)} Additional rejection factor as a function of momentum for pion tracks
faking electrons for the same DV cut values as measured within this analysis
(for illustration only).}
\end{figure}

Further suppression of pions mis-identified as
electrons within the above conditions is obtained by using a discriminating 
variable (DV) which is a linear combination of three quantities related to 
shower properties ($E/p$, radial shower width, and energy-weighted 
track-cluster
distance at the LKr front face), and is almost momentum independent. 
This variable was developed as described in Ref. \cite{ke410} and  was 
trained on dedicated track samples to be close to 1 for electron tracks and 
close to 0 for pion tracks misidentified as electron tracks.
In the signal selection, its value is 
required to be larger than 0.9 for the electron track candidate. When taking
into account the electron momentum spectrum, the resulting efficiency is above
96\% (Fig. \ref{fig:lda}).

\begin{figure}[htp]
\begin{center}
\begin{picture}(160,150)
\put(-148,-50){\epsfxsize78mm\epsfbox{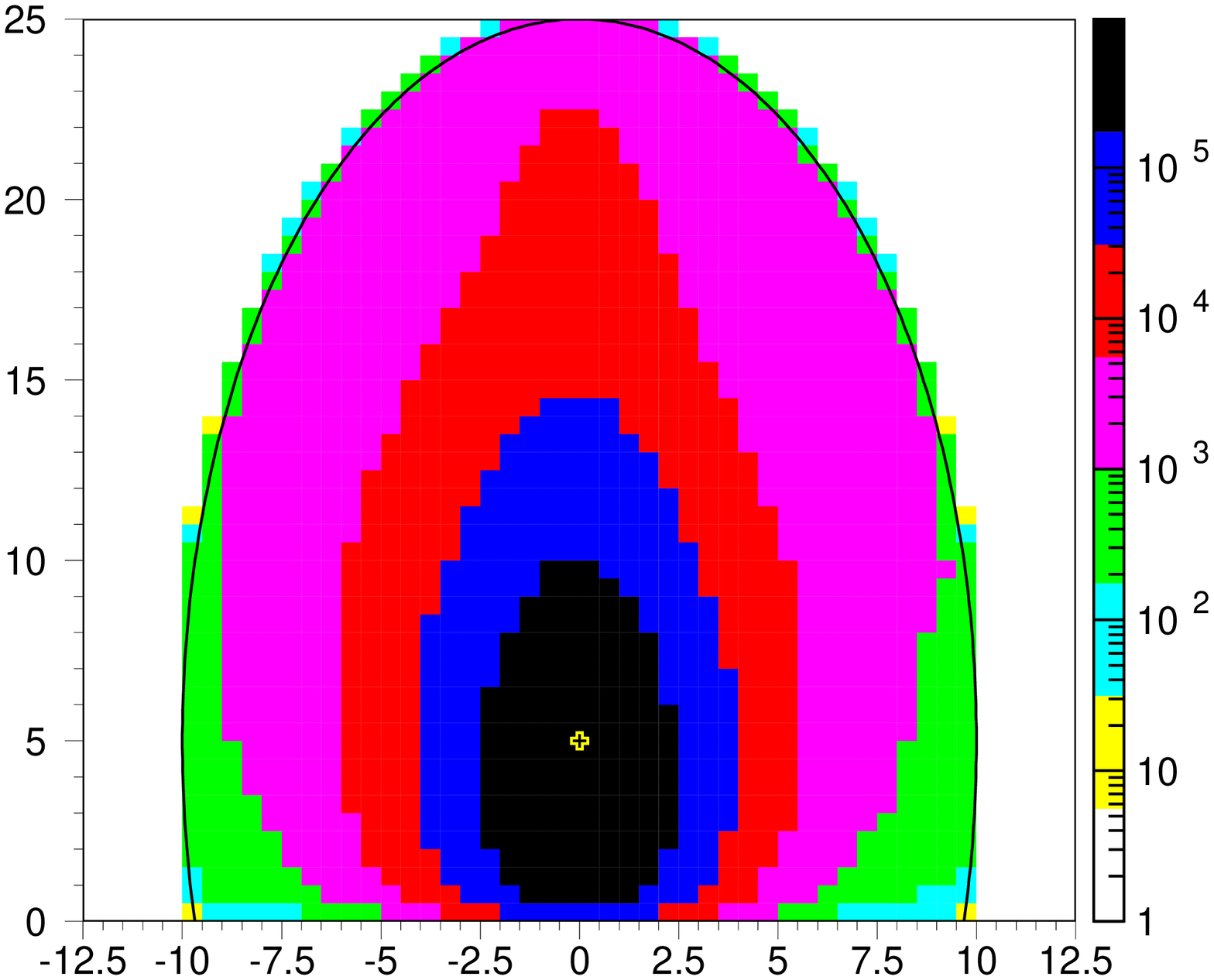}}
\put(82,-50){\epsfxsize78mm\epsfbox{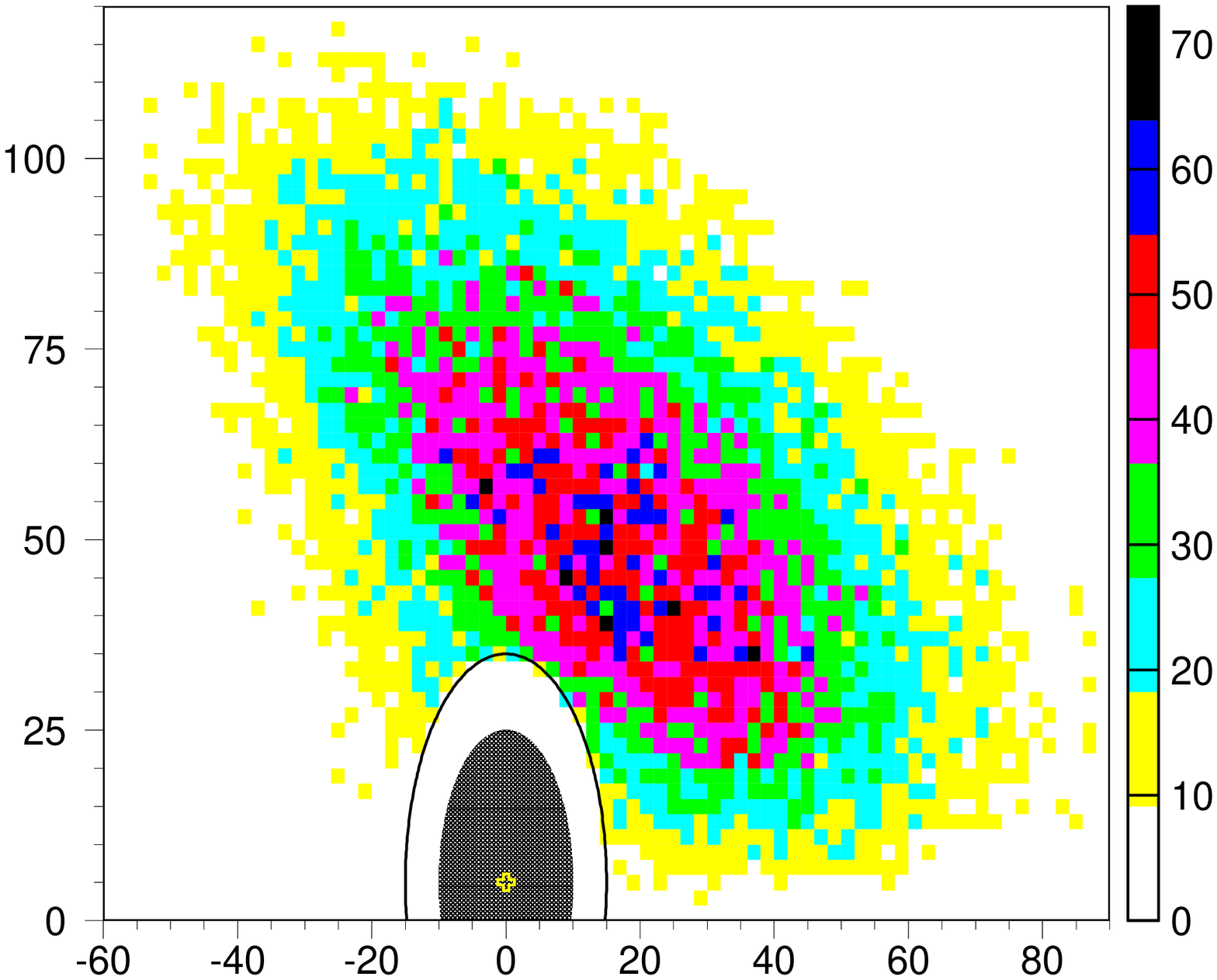}}
\put(9,118){$\KTPO$ Data}
\put(239,118){$\KEQO$ Data}
\put(-25,-52){$M_{3\pi} - M_K ~(\MEVcc)$}
\put(205,-52){$M_{3\pi} - M_K ~(\MEVcc)$}
\put(-135,143){$p_t ~(\MEVc)$}
\put(95,143){$p_t ~(\MEVc)$}
\put(-118,118){\bf \large (a)}
\put(112,118){\bf \large (b)}
\end{picture}
\vspace{10mm}
\end{center}
\caption{Reconstructed ($M_{3\pi} , p_t$) plane for the normalization
{\bf (a)} and signal {\bf (b)} candidates (note the 
different color scales). The left plot is a zoom inside the smaller ellipse 
which defines the normalization sample. 
Crosses correspond to the ellipse centers ($M_{3\pi} = M_K, ~p_t = 5~\MEVc)$.
\label{fig:mkpt}}
\end{figure}

{\bf Normalization sample.} In the plane $(M_{3\pi},p_t)$, the $\KTPO$ sample 
is selected by the requirement to be inside an ellipse centered on the nominal 
kaon mass and a $p_t$ value of $5 ~\MEVc$, with semi-axes $10~\MEVcc$ and $20~\MEVc$, 
respectively, thus requiring fully reconstructed $\KTPO$ three-body decay (Fig.
\ref{fig:mkpt}a). 

The parent kaon momentum $|\sum\vec p_i|$ is required to be reconstructed  
between 54 and 66 $\GEVc$ and the vertex is required to be composed of a pair 
of $\PIo$ candidates and a {\it pion} candidate. A total of $93.54 \times 10^6$
candidates satisfies the above criteria.

{\bf Signal sample.} In the plane $(M_{3\pi},p_t)$, the $\KEQO$ sample is 
obtained requiring candidates to be outside an ellipse centered on the nominal 
kaon mass and a $p_t$ value of $5 ~\MEVc$, with semi-axes $15~\MEVcc$ and 
$30~\MEVc$, respectively, allowing any $p_t$ value for the undetected neutrino 
and 
rejecting $\CUSP$ fully reconstructed three-body decays (Fig. \ref{fig:mkpt}b).

The reconstructed parent kaon momentum under the $\KEQ$ hypothesis is required 
to be in the fiducial range between 54 and 66 $\GEVc$
and the vertex is required to be composed of a pair of $\PIo$ candidates and an
{\it electron} candidate. 

The neutrino momentum vector is then defined as the missing momentum in the
equation $\vec{p}_{\nu} = \vec{p}_{K} - \vec{p}_{e} - \vec{p}_{\pi^0 _1}  - \vec{p}_{\pi^0 _2}$ and is used to compute
the invariant mass of the electron-neutrino system, which is required to be 
smaller than the maximum kinematic value of 0.25 $\GEVcc$. 
A total sample of 65210 candidates is selected.

\section{Background estimate}
\label{sec:bkg}
The $\KTPO$ decay is the most significant background source contributing to the
 $\KEQO$ signal. It contributes either via the decay in flight of the charged
 pion ($\PMP \to e^\pm\nu$, genuine electron) or mis-identification of the 
pion as an electron (fake electron).
In the genuine electron case, only pion decays occurring close to the parent 
kaon decay vertex or leading to a
forward electron and thus consistent with the  neutral vertex and 
($M_{3\pi}, p_t$) requirements may satisfy the signal selection.
Another accidental source of background to both signal and normalization 
samples occurs when an additional track or photon combines with another kaon 
decay (for example $\mathrm{K}^{\pm}_{2\pi\gamma}$ or
$\mathrm{K}^{\pm}_{\mathrm{e}3\gamma}$) and forms a fake $\KTPO$ or $\KEQO$ 
final state, or replaces a real track or photon in a $\KTPO$ or $\KEQO$ decay.

\begin{figure}[htb]
\begin{center}
\begin{picture}(160,70)
\put(-140,-136){\epsfxsize80mm\epsfbox{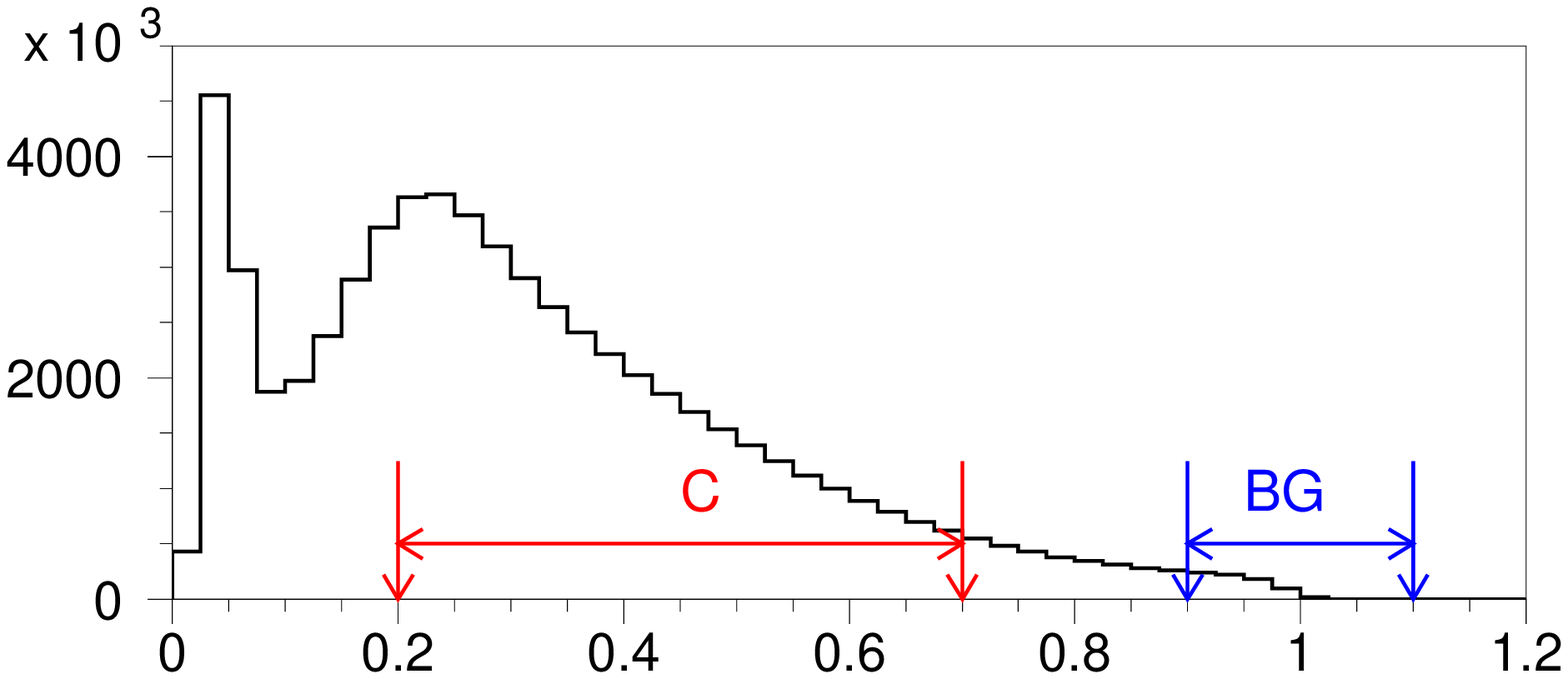}}
\put(45,40){\bf \large (a)}
\put(20,40){$\KTPO$}
\put(50,-40){$E/p$}
\put(90,-136){\epsfxsize80mm\epsfbox{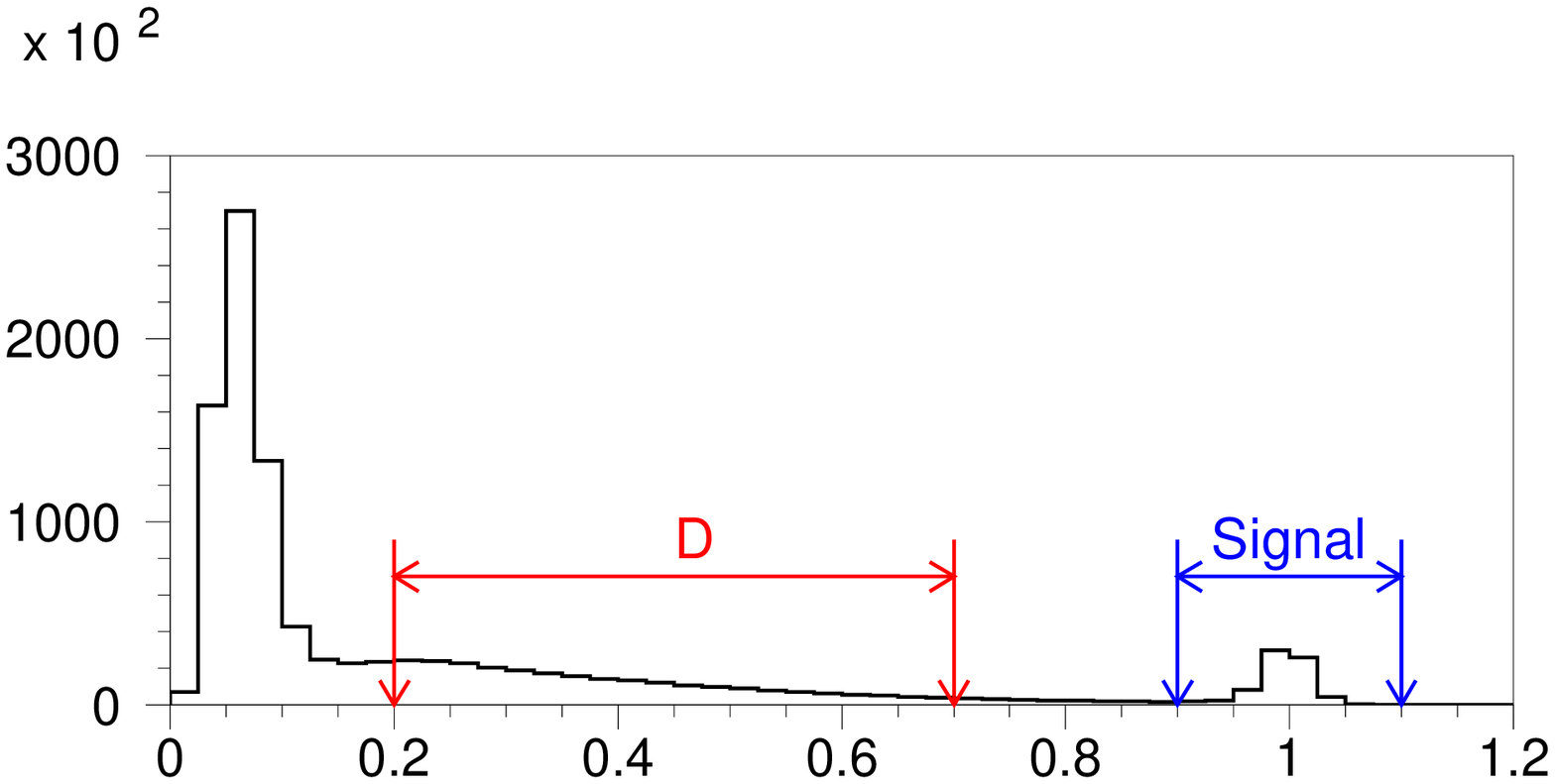}}
\put(275,40){\bf \large (b)}
\put(250,40){$\KEQO$}
\put(280,-40){$E/p$}
\end{picture}
\end{center}
\caption{\label{fig:fak}
Distribution of the $E/p$ ratio before electron-identification criteria are 
applied. Three components (in different proportions) are visible: muons with 
low $E/p$ values, electrons at $E/p$ values close to 1 and pions in between. 
Regions of interest are:
{\bf (a)} control C and BG regions in the $\KTPO$ sample;
{\bf (b)} control D and Signal regions in the $\KEQO$ sample.
}
\end{figure}

The remaining fake-electron background after the DV requirement can be studied 
from a subset of the normalization data sample whose selection does not rely on
any LKr requirement. Two subsamples having a track pointing to the LKr fiducial
acceptance, associated with an in-time energy cluster and away from the closest
{\it inactive} cell by more than 2 cm are considered: the control sample C 
($N_C$ events) with $E/p$ between 0.2 and 0.7, and the background sample BG 
($N_{BG}$ events) with full electron-identification requirement ($E/p$ 
between 0.9 and 1.1 and DV above 0.9). The 
ratio $N_{BG}/N_{C}$ characterizes the fraction of fake-electrons kept after 
electron-identification. This fraction has a weak dependence on the track 
momentum and is typically a few $10^{-3}$.
In the signal sample, a similar control sample D ($N_D$ events) with the same
$E/p$ range is defined 
before electron-ID
requirements are applied. The background from fake-electron tracks in the 
signal region is obtained as  $N_D \times N_{BG} / N_C$ and amounts to 
$425 \pm 2$ events (0.65\% relative to signal candidates). 
This is illustrated in Fig. \ref{fig:fak}.

\begin{figure}[htb]
\begin{center}
\begin{picture}(160,60)
\put(-140,-130){\epsfxsize80mm\epsfbox{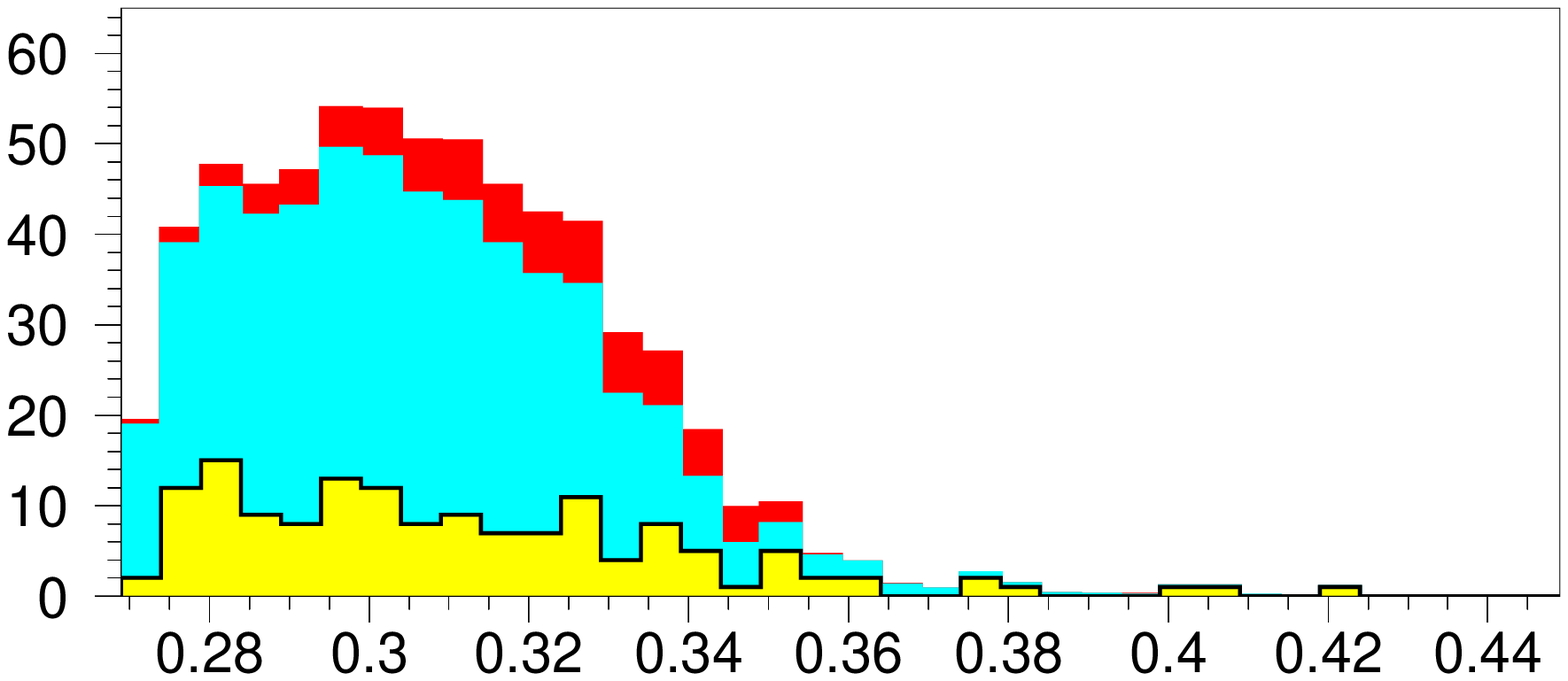}}
\put(10,-40){$\MPP ~(\GEVcc)$}
\put(-110,45){\bf \large (a)}
\put(90,-130){\epsfxsize80mm\epsfbox{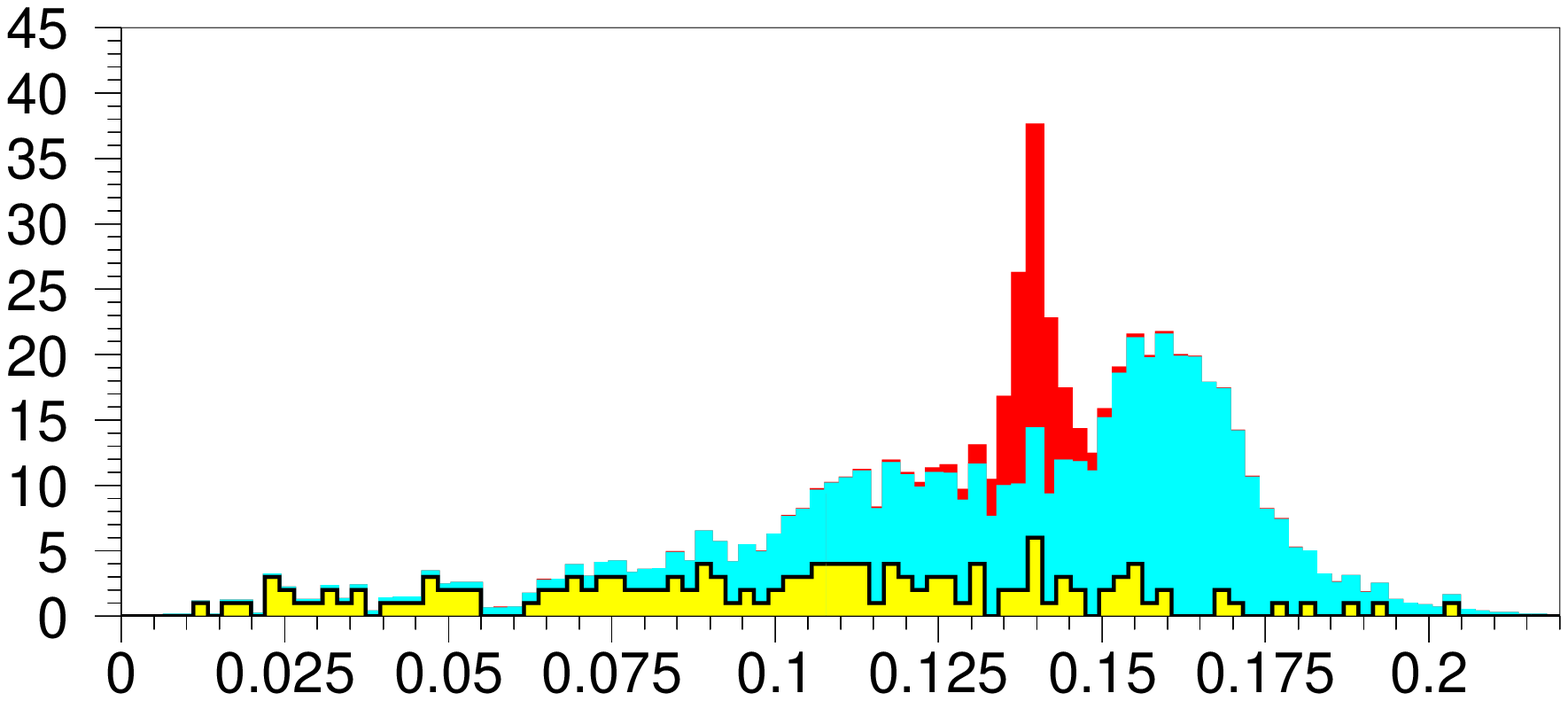}}
\put(240,-40){$\MEN ~(\GEVcc)$}
\put(280,45){\bf \large (b)}
\put(-85,-105){\epsfxsize80mm\epsfbox{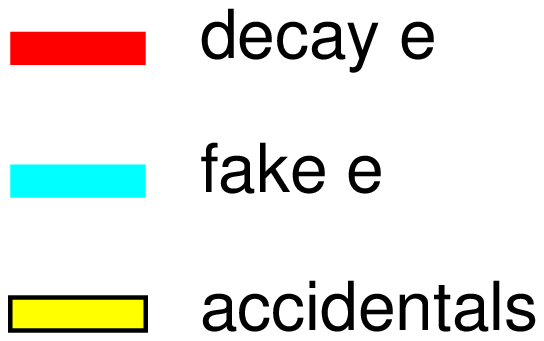}}
\put(40,-105){\epsfxsize80mm\epsfbox{capfig4.eps}}
\end{picture}
\end{center}
\caption{\label{fig:bkg}
 Cumulative distribution of $\MPP$ {\bf (a)} and $\MEN$ {\bf (b)} variables for
background contributions to the $\KEQO$ candidates
from accidentals, fake electrons and decay electrons.
While accidental background is uniformly 
distributed, fake and decay electrons are concentrated at low $\MPP$ mass
values and $\MEN$ mass values close to $m_{\pi^+}$.
}
\end{figure}

The contribution of genuine electrons from pion decay ($\PMP \to e^\pm\nu$) is
strongly suppressed because of 
its small branching ratio 
$(1.23 \times 10^{-4})$ combined with the pion decay probability before the LKr
($\sim 10\%$). To get a large enough sample, a dedicated $\KTPO$ simulation 
where the charged pion is only decaying to $e \nu$ has been studied. 
This contribution to the signal candidate sample amounts
to $79 \pm 1$ events (0.12\% relative to signal candidates). 
For this background, the 
reconstructed invariant mass of the $e \nu$ system peaks, as expected, at the 
charged pion mass smeared by detector resolution (Fig. \ref{fig:bkg}).

Accidental background has been studied in both signal and normalization samples
by loosening the timing cuts either between the four photons or between the
track and the four photons. The number of candidates selected in the 
side bands of the time distributions has been extrapolated to the selection
region. 
The accidental contribution is estimated to be $231078 \pm 481$ events in the
normalization sample and $146 \pm 12$ events in the signal sample, 
corresponding to relative contributions of 
$(2.470 \pm 0.005)\times 10^{-3}$ 
and $(2.240 \pm 0.202)\times 10^{-3}$, respectively.

The distributions of the two invariant masses $\MPP$ and $\MEN$ built from the
three background sources with appropriate scaling are displayed in Fig. 
\ref{fig:bkg}. The relative background contribution to the selected $\KEQO$ 
sample is estimated to be $(1.00 \pm 0.02)\%$, dominated by the fake electron 
component from $\KTPO$ decays.

\section{\boldmath Theoretical formalism}
\begin{figure}[htp]
\vspace{-7cm}\hspace{2cm}
\mbox{\epsfig{file=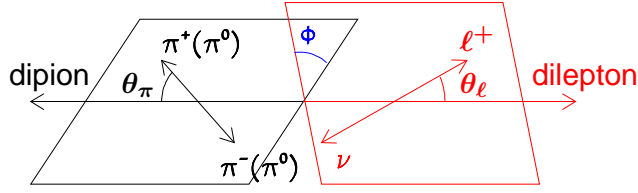 ,width=12cm}}
\put(-110,115){\textcolor{red}{\boldmath $\ell^{+}$}}
\put(-110,99){\textcolor{red}{\boldmath $\theta_{\ell}$}}
\put(-238,99){\boldmath $\theta_{\pi}$}
\vspace{-2cm}
\caption{Sketch of the $\KLQ$ decay in the kaon rest frame showing the 
definitions of $\theta$ and $\phi$ angles within and between the dipion and 
dilepton planes.
\label{fig:sketch}}
\end{figure}

The differential rate of the 
$\KLQ$ decay ($\ell = \mu$, e) of a $\KPL$ is described by five kinematic 
variables (historically called Cabibbo-Maksymowicz variables~\cite{cabmak}) as 
shown in Fig. \ref{fig:sketch}:

- $\SP = M_{\pi\pi}^2$, the square of the dipion invariant mass,

- $\SL = M_{\ell\nu}^2$, the square of the dilepton invariant mass,

- $\theta_{\pi}$, the angle of the $\PPI$ ($\PIo$) in the dipion rest frame 
with respect to the direction of flight of the dipion in the kaon rest frame,

- $\theta_{\ell}$, the angle of the $\ell^{+}$ in the dilepton rest frame
with respect to the direction of flight of the dilepton in the kaon rest frame,

- $\phi$, the azimuthal angle between the dipion and dilepton planes in the kaon
 rest frame.

The decay amplitude is written as the product of the weak current of the 
leptonic part and the (V -- A) current of the hadronic part:

\begin{displaymath}
 \displaystyle\frac{ G_{F}}{\sqrt{2}} V^{*}_{us}  ~\bar{u}_{\nu}\gamma_{\lambda}
(1-\gamma_{5}) v_{\ell}~\langle\POP \POM | V^{\lambda}- A^{\lambda} | \KPL\rangle,  \rm{~where} \nonumber
\end{displaymath}
\begin{eqnarray}\label{eq:fghr}
\langle\POP \POM| A^{\lambda}|\KPL\rangle  =&\displaystyle\frac{-i}{m_{K}} \left(F(\PBO_{\POP} + \PBO_{\POM})^{\lambda} 
+  ~G(\PBO_{\POP} - \PBO_{\POM})^{\lambda} + R(\PBO_{\displaystyle\ell} + \PBO_{\displaystyle\nu})^{\lambda}\right)
 \rm{ ~~~and} \nonumber \\
\langle\POP \POM| V^{\lambda}|\KPL\rangle  = &\displaystyle \frac{-H}{m_{K}^3}\epsilon^{\lambda\mu\rho\sigma} (\PBO_{\POP} + \PBO_{\POM} + \PBO_{\displaystyle\ell} + \PBO_{\displaystyle\nu} )_{\mu} \hspace{51mm} \nonumber \\
& \hspace{30mm} \times ~
(\PBO_{\POP} + \PBO_{\POM})_{\rho} (\PBO_{\POP} - \PBO_{\POM})_{\sigma} .
\end{eqnarray}

In the above expressions, $\PBO$ refers to the four-momentum of the final state
particles, $F, G, R$ are three axial-vector and $H$ one vector complex form
factors with the convention $\epsilon^{0123} = 1$. Note that $F, R$ are 
multiplied by terms symmetric with respect to the exchange of the two pions, 
while $G, H$ are multiplied by terms antisymmetric with respect to this same 
exchange.

The decay probability summed over lepton spins can be expressed as:
\begin{equation}\label{eq:dg5}
d^{5}\Gamma = \displaystyle\frac{G_{F}^{2}  |V_{us}|^{2}}{2 (4 \pi)^{6} m_{K}^5} ~\rho(\SP,\SL) ~J_{5}(\SP,\SL,\THP,\THL,\phi)  ~d \SP ~d \SL ~d \CTP ~d\CTL ~d\phi,
\end{equation}

\noindent where $ \rho(\SP,\SL) = X \sigma_{\pi}\left( 1 - z_{\ell} \right)$ 
is the phase space factor, with
 $ ~X = \HLF \lambda^{1/2}(m_{K}^2,\SP,\SL),\\
   ~\sigma_{\pi} = (1 -4m _{\pi}^{2}/\SP)^{1/2},
~\ZL = m_{\ell}^2 / \SL,$ and
$\lambda(a,b,c) = a^2 + b^2 + c^2 - 2 ( ab + ac + bc). $

The function $J_5$, displaying the angular dependencies on $\THL$ and $\phi$,
reads~\cite{paist,BCG}:
\begin{eqnarray}\label{eq:j3}
J_{5} = & 2 (1 - \ZL)  (I_{1} + I_{2}~\CTTL + I_{3}~\SSTL \cdot\cos2\phi
 + I_{4}~\STTL \cdot \CPH  + I_{5} ~\STL \cdot\CPH  \nonumber \\
          &                 + I_{6}~\CTL + I_{7}~\STL\cdot\SPH
 + I_{8}~\STTL \cdot\SPH + I_{9}~\SSTL \cdot\sin2\phi), 
\end{eqnarray}
where 

$
\begin{array}{rl}
I_{1} &= \QRT\left( (1 + \ZL)|F_{1}|^{2} + \HLF(3+ \ZL)(|F_{2}|^{2} + |F_{3}|^{2})~\SSTP + 2 \ZL|F_{4}|^{2}\right) ,\vspace{2mm}\\
I_{2} &= -\QRT (1 - \ZL)\left( |F_{1}|^{2} -\HLF (|F_{2}|^{2} + |F_{3}|^{2})~\SSTP \right), \vspace{2mm}\\
I_{3} &= -\QRT (1 - \ZL)\left( |F_{2}|^{2} - |F_{3}|^{2} \right)~\SSTP  ,\vspace{2mm}\\
I_{4} &= \HLF (1 - \ZL) Re(F_{1}^{*}F_{2})~\STP,\vspace{2mm}\\
I_{5} &= - \left(Re(F_{1}^{*}F_{3}) + \ZL~Re(F_{4}^{*}F_{2}) \right)~\STP,\vspace{2mm}\\
I_{6} &= - \left(Re(F_{2}^{*}F_{3})~\SSTP - \ZL~Re(F_{1}^{*}F_{4}) \right),\vspace{2mm}\\
I_{7} &= - \left(Im(F_{1}^{*}F_{2}) + \ZL~Im(F_{4}^{*}F_{3}) \right)~\STP,\vspace{2mm}\\
I_{8} &= \HLF(1 - \ZL) ~Im(F_{1}^{*}F_{3} )~\STP,\vspace{2mm}\\
I_{9} &= -\HLF(1 - \ZL)~ Im(F_{2}^{*}F_{3})~\SSTP .
\end{array}
$

\vspace{5mm}
The $I_1$ to $I_9$ expressions carry the dependence on ($\SP,~\SL,~\THP$) using
the form factors ($F_{i}, i=1,4$), combinations of the
complex hadronic form factors $F,~G,~R,~H$ defined in Eq. (\ref{eq:fghr}).
  
In $\KEQ$ decays, the electron mass can be neglected $(\ZL = 0)$ and the
terms $( 1 \pm  \ZL)$ become unity. One should also note that the form
factor $F_{4}$ is always multiplied by $\ZL$ and thus does not contribute
to the full expression. 

In the case of the neutral pion mode, there is no unambiguous definition
of the $\theta_{\pi}$ angle as the two $\PIo$ cannot be distinguished. 
The form factors $F_{2} =\sigma_{\pi} ( \SP \SE )^{1/2} G$ and
$F_{3} =\sigma_{\pi} X \left( \SP \SE \right)^{1/2} H / m_{K}^2 $ 
are related to the 
$G,H$ form factors of the decay amplitude, antisymmetric in the exchange of the
two pions and therefore of null values.

With this simplification, there is a single complex hadronic form factor 
$F_1 =  X F + \HLF \sigma_{\pi} (m_{K}^2 - \SP - \SE) ~\CTP ~G$ in the 
expression of $J_5$ which then reads $F_1 =  X F $, symmetric in the 
exchange of the two pions.
At leading order, only the S-wave component of the partial wave expansion 
contributes ($F \equiv m_{K}^2 F_s$ where $F_s$ is dimensionless).

The integration over the variables $\CTP$ and $\phi$ is trivial and 
Eqs. (\ref{eq:dg5}, \ref{eq:j3}) become:
\begin{eqnarray}\label{eq:dg3}
d^{3}\Gamma &= \displaystyle\frac{G_{F}^{2}  |V_{us}|^{2}}{4 (4 \pi)^{5} m_{K}^{5} } ~\rho(\SP,\SE)~J_{3}(\SP,\SE,\CTE) ~d\SP ~d\SE ~d\CTE, \nonumber \\
J_{3}& = \HLF |X  F|^2 (1 - \CTTE) = m_{K}^4 |X  F_{s}|^2 \SSTE . \phantom{~d\SP ~d\SE ~d\CTE}
\end{eqnarray}

The differential rate depends on a single form factor $F_s$ whose variation 
with (\SP,~\SE) is unknown and will be studied.

\section{Acceptance calculation}
\label{sec:accep}
A detailed GEANT3-based~\cite{geant} Monte Carlo (MC) simulation is used to
compute the acceptance for signal and normalization channels.
It includes full detector geometry and material description,
stray magnetic fields, DCH local inefficiencies and misalignment, LKr local
inefficiencies, accurate simulation of the kaon beam line (reproducing the 
observed flux ratio $\KPL / \KMI \sim 1.8$) and time variations
of the above throughout the running period.
This simulation is used to perform two time-weighted MC productions, 
$10^8$ generated decays each, large enough to obtain the acceptances with a 
relative precision of few $10^{-4}$.  

The signal channel $\KEQO$ is generated according to Eq. (\ref{eq:dg3}) 
including a constant $F_s$ form factor. It can be reweighted according to
another description of the form factor as obtained for example in 
Ref.~\cite{ke410} or in this analysis.
The chosen form factor value is then propagated to the acceptance calculation 
by means of the same reweighting procedure.  Going from a constant 
form factor value to the energy dependent value measured in Ref.~\cite{ke410}, 
the relative signal acceptance change is $- 1 \%$.

\begin{figure}[htp]
\begin{center}
\begin{picture}(160,150)
\put(-140,-50){\epsfxsize80mm\epsfbox{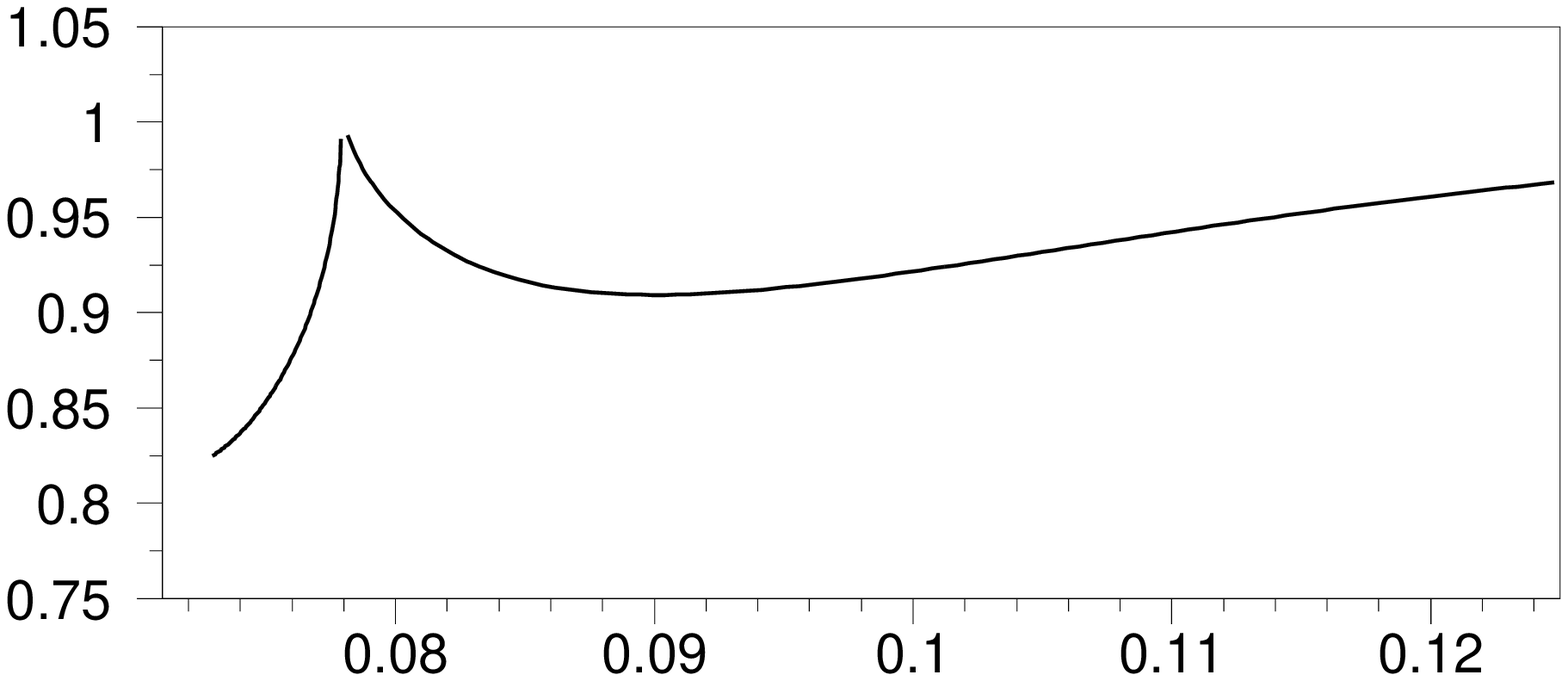}}
\put(-15,40){$M_{\pi^0 \pi^0}^2 ~(\GEVcc)^2$}
\put(-110,125){\bf \large (a)}
\put(45,125){$\KTPO$}
\put(90,-50){\epsfxsize80mm\epsfbox{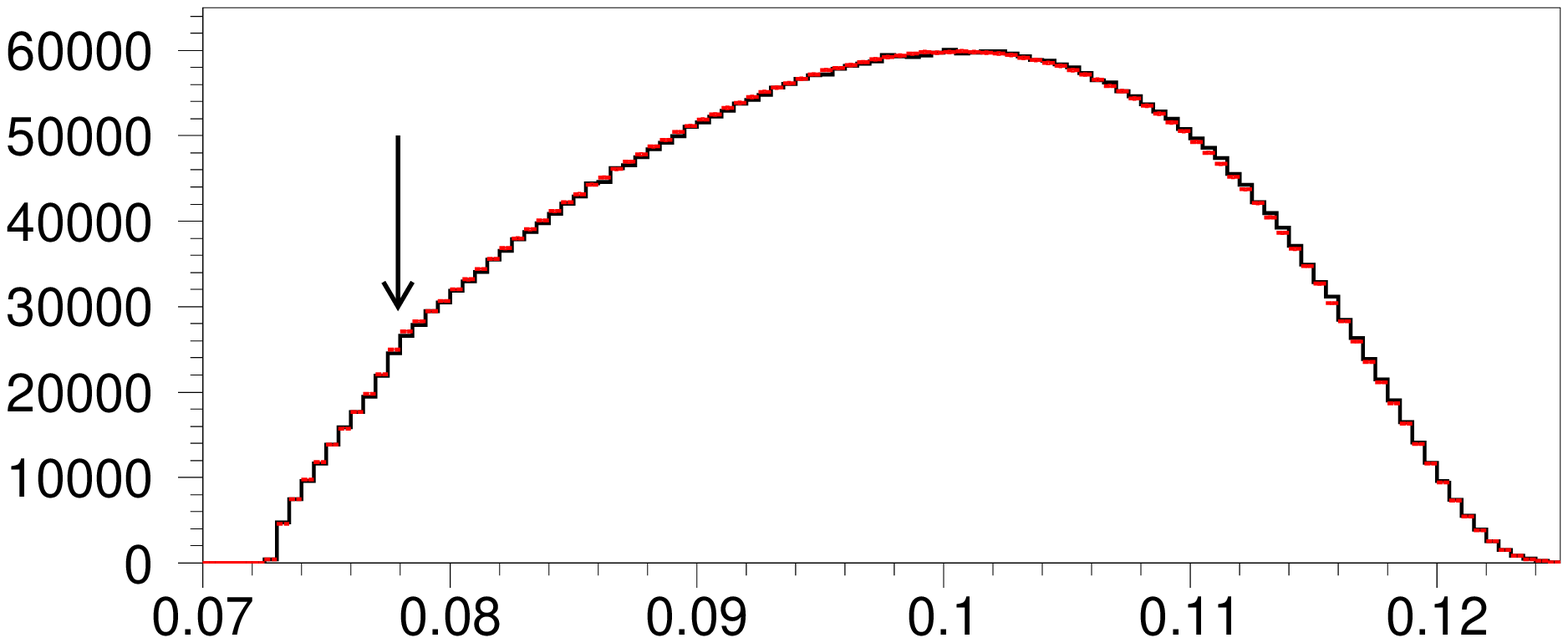}}
\put(210,40){$M_{\pi^0 \pi^0}^2 ~(\GEVcc)^2$}
\put(120,125){\bf \large (b)}
\put(275,125){$\KTPO$}
\end{picture}
\end{center}
\vspace{-25mm}
\caption{{\bf (a)} Cusp-like 
modification to the $\PIo\PIo$ invariant mass squared as
introduced in the simulation, normalized to the classical series expansion in
$M_{\pi^0 \pi^0}^2$. 
{\bf (b)} Distributions of the $\PIo\PIo$ invariant mass squared for 
reconstructed data and simulated events. The arrow points to the  
$4 m_{\pi^+}^2$  value.
\label{fig:k3p}}
\end{figure}

The normalization channel $\KTPO$ is well understood in terms of simulation,
being of primary physics interest to NA48/2~\cite{cusp}. The most precise 
description of the decay amplitude has been implemented. This description 
corresponds to an empirical parameterization of the data~\cite{empir} which
includes the cusp-like shape of the $\PIo\PIo$ invariant mass squared at the 
$4 m_{\pi^+}^2$ threshold and $\pi \pi$ bound states
(Fig.~\ref{fig:k3p}).

Depending on the data taking conditions, the relative acceptance variation can 
be as large as 5\% 
for both signal and normalization channels  due to the faulty HOD slabs but the
 ratio $A_{n} / A_{s}$ 
stays within $\pm 0.4\%$ of its average value.

The same selection and reconstruction as described in Section~\ref{sec:selevt}
are applied to the simulated events except for the trigger and timing 
requirements.
Particle identification cuts related to the LKr response are replaced by
momentum-dependent efficiencies, obtained from data in pure samples of electron
tracks (Fig. \ref{fig:lda}a).

Real photon emission using {\tt PHOTOS} 2.15~\cite{photos} is included in both
$\KEQO$ and $\KTPO$ simulations. It distorts the original $M_{e\nu}$ 
distribution and consequently modifies the overall acceptance.
A dedicated study with and without photon emission was performed on a subset 
of the simulation sample. The $\KTPO$ acceptance is 
unaffected while a relative acceptance change of $-2\%$ is observed for $\KEQO$
when real photon emission is implemented.
Details of the photon emission modeling will be discussed together with other
systematic uncertainties. 

\begin{figure}[htp]
\begin{center}
\begin{picture}(160,150)
\put(-150,-50){\epsfxsize80mm\epsfbox{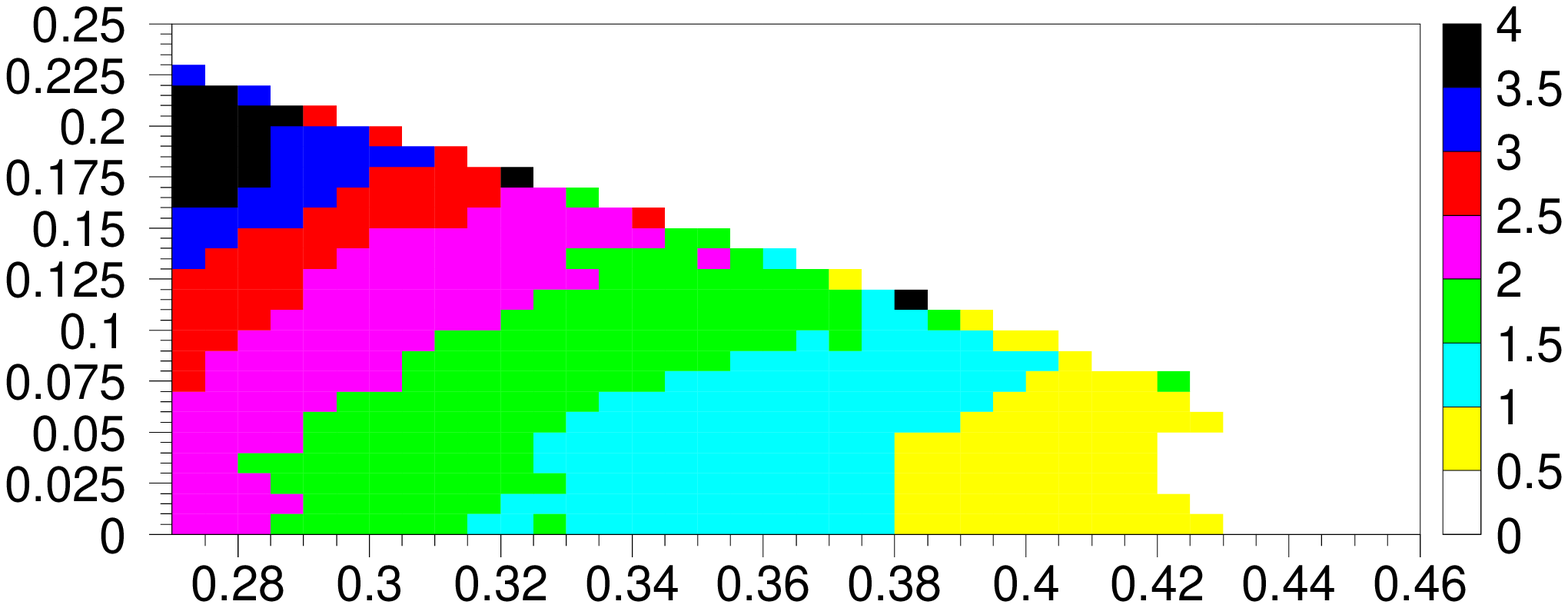}}
\put(5,40){$\MPP ~(\GEVcc)$}
\put(-140,145){$\MEN ~(\GEVcc)$}
\put(49,145){A($\%$)}
\put(22,125){$\KEQO$ {\bf \large (a)} }
\put(80,-50){\epsfxsize80mm\epsfbox{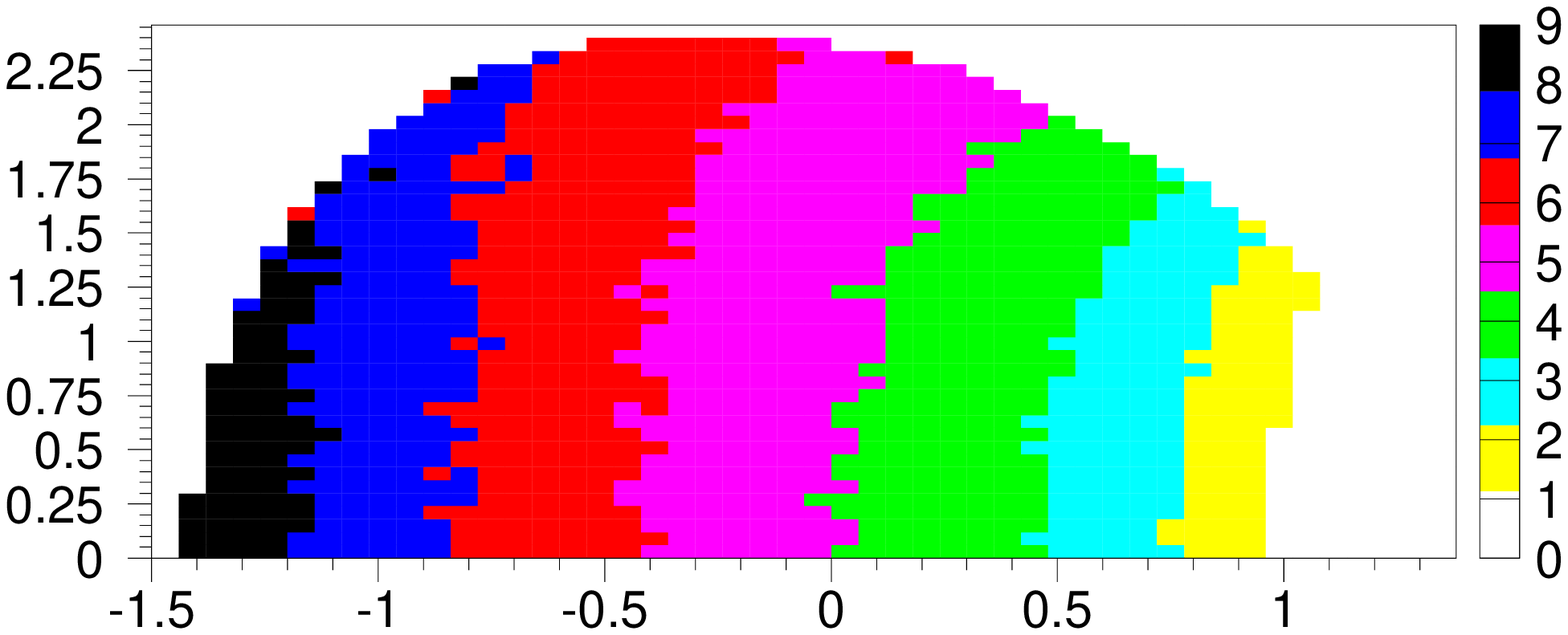}}
\put(280,40){$u$}
\put(100,145){$|v|$}
\put(279,145){A($\%$)}
\put(250,125){$\KTPO$ {\bf \large (b)}}
\end{picture}
\end{center}
\vspace{-25mm}
\caption{Acceptance of {\bf (a)} $\KEQO$ candidates in the plane ($\MPP,\MEN$) 
and of {\bf (b)} $\KTPO$ candidates in the plane ($u, |v|$). The dimensionless 
variables 
($u, |v|$) are defined as $u = (M_{\pi^0 \pi^{0}}^2 - s_0 ) / m_{\pi^{+}}^2$ 
and 
$|v| = |M_{\pi^+ \pi^{0}_1}^2 - M_{\pi^+ \pi^{0}_2}^2 | / m_{\pi^{+}}^2$  
with $s_0 = (m_{K^{+}}^2 + 2 m_{\pi^{0}}^2  + m_{\pi^{+}}^2)/3$.
The notation $\pi^{0}_1 ,\pi^{0}_2$ is only used to distinguish one $\PIo$ from
the other but has no particular meaning beyond this.
\label{fig:acc}}
\end{figure}

The acceptance values, 
averaged on both kaon charges and over the data-taking periods, are 
$A_{s} = (1.926 \pm 0.001) \%$  and $A_{n} = (4.052 \pm 0.002) \%$.
The $A_{s}$ variations from 0 to about 4\% ($A_{n}$ from 0 to about 9\%) across
the Dalitz plot are shown in Fig. \ref{fig:acc}.

\section{Trigger efficiency}
\label{sec:trig}
Both signal and normalization modes are recorded concurrently with the same
trigger logic. Downscaled minimum bias control triggers 
are used to measure the efficiency of the main trigger 
channels.
Hardware changes to the trigger conditions were introduced during 
data taking following improvements in detector and readout electronics
performance. 
As a consequence, trigger effects have been studied separately for data samples
taken during ten periods of stable trigger conditions.
Details of the trigger efficiency for normalization events are given in Refs.
\cite{agcomb,ag0}.
As described in Section \ref{sec:beamdet}, $\KTPO$ and $\KEQO$ events were
recorded by a first level trigger using signals from HOD
(Q1) and LKr (NUT), followed by a second level trigger using DCH 
information (MBX).
Using event samples recorded with downscaled control triggers, and selecting
$\KTPO$ and $\KEQO$ decays as described in Section \ref{sec:selevt}, it is
possible to measure separately two efficiencies:

- the efficiency of the NUT trigger using a sample recorded by the 
Q1$\cdot$MBX trigger;

- the efficiency of the Q1$\cdot$MBX trigger using a sample recorded by 
the NUT trigger.

These two efficiencies rely on different detector information and are 
statistically independent. 
They are multiplied to obtain the overall trigger efficiency of each 
subsample for both signal and normalization channels.

{\bf NUT trigger efficiency.} In the $\KTPO$ selection, the inefficiency 
is measured to be 0.5\% (most of 2003), 3\% (end of 2003 and beginning of
2004) and then $3 \times 10^{-4}$ until the end of 2004. Because of the extra 
LKr energy deposit from the electron, this inefficiency is even smaller in the 
$\KEQO$ selection than in the  $\KTPO$ selection.
In each data taking period, the control trigger sample is large enough to 
determine the efficiency with an excellent precision ${\cal O}(10^{-4})$ for 
the normalization sample and a precision better than $5 \times 10^{-3}$ in the 
signal sample.

{\bf Q1$\cdot$MBX trigger efficiency.} The inefficiency suffers from somewhat 
large variations with data taking conditions, ranging from 3\% to 7\% due to
local DCH inefficiencies. Control samples
are large enough in the $\KTPO$ selection to determine the efficiency within a
few $10^{-4}$ precision. In the signal selection, there are too few control 
triggers in 2003 to ensure a precise enough efficiency measurement. As the 
uniformity of the Q1 trigger part is ensured by HOD geometrical fiducial cuts 
in the selection, the lack of statistics is overcome by taking advantage of 
the realistic simulation code of the MBX algorithm that proves to reproduce 
accurately the 
efficiency variations in the $\KTPO$ selection, as measured from the data. 
The MBX efficiency for the signal mode is therefore obtained from the MBX 
simulation.
The Q1$\cdot$MBX efficiency values are in very good agreement with the 
measured values but obtained with improved precision.

The statistical average of the Q1$\cdot$NUT$\cdot$MBX trigger efficiency
over the ten independent samples is $(96.06 \pm 0.03)\%$ for the $\KEQO$ 
selection and $97.42 \%$ with a negligible error for the $\KTPO$ selection.

\section{Form factor measurement}
\label{sec:ff}
\subsection{Measurement method}
The form factor study requires a sample free of large radiative effects which
can pollute the original kaon decay amplitude. An extra cut is applied in
the signal selection (Section \ref{sec:selevt}), rejecting events where an 
additional photonic energy deposit is identified with at least $3~\GEV$ 
energy, in-time with the signal candidate track and photons, and away by more 
than 15 cm from the track impact at LKr and 10 cm from each of the four photons
forming the two $\PIo$ candidates. This reduces the number of selected signal
candidates from 65210 to 65073 and the estimated number of background events 
from 650 to 641.

The event density in the $(\SP ,\SE)$ plane, also called the Dalitz plot, is 
proportional to $F_s ^2$ as shown in Eq. (\ref{eq:dg3}). The number of events
in the 
$(\SP ,\SE)$ plane and the projected distributions along the two variables are 
displayed in Fig.~\ref{fig:spise}. The Dalitz plot density is compared, after
background subtraction, to the density obtained from the simulation where 
kinematics, acceptance, resolution, trigger efficiency, and radiative effects 
are taken into account.

To analyze the data as a single sample while reproducing the variation of data 
taking conditions in the simulation as closely as possible, the simulated 
sample should reflect:\\
- the time dependence of the number of kaon decays in each data sample;\\
- the relative variation of the trigger efficiency across the Dalitz plot;\\  
- the measured Q1$\cdot$NUT$\cdot$MBX trigger efficiency in each subsample.

To that purpose, the integrated number of kaon decays in each subsample is 
obtained  from
the number of observed normalization candidates corrected for the 
selection acceptance, the trigger efficiency and the known branching ratios 
\cite{pdg}.
The numbers of $\KEQO$ signal events generated for each subsample is the same
up to an arbitrary scale factor reflecting the fraction of the
total number of kaon decays generated. A fine tuning of the generated 
subsample sizes results in applying similar weights to all subsamples, in 
the range 0.9 to 1.

No particular pattern is observed in the Dalitz plot for the inefficient 
NUT triggers.
The Q1 trigger efficiency is known to be very high $( > 99.75\%$ as measured in
other studies \cite{agcomb}) and uniform 
once the local inefficient areas have been excluded by the event selection. 
The dependence of the MBX trigger efficiency on $\SP$ can be studied from
$\KTPO$ data control triggers and simulated samples. Because of different local
DCH inefficiencies, different subsamples may have non-identical variations. 
However the variations observed in the data are well reproduced in the 
simulation within a $\pm1\%$ relative accuracy. This justifies the usage of 
the simulation code as being realistic also for the $\KEQO$ signal. 

Once local variations have been considered, a fine tuning of the overall 
trigger efficiency of each simulated subsample is achieved by applying 
weights with values between 0.98 and 1.

\subsection{Fitting procedure and results}
\begin{figure}[htp]
\begin{center}
\begin{picture}(160,160)
\put(-148,-45){\epsfxsize80mm\epsfbox{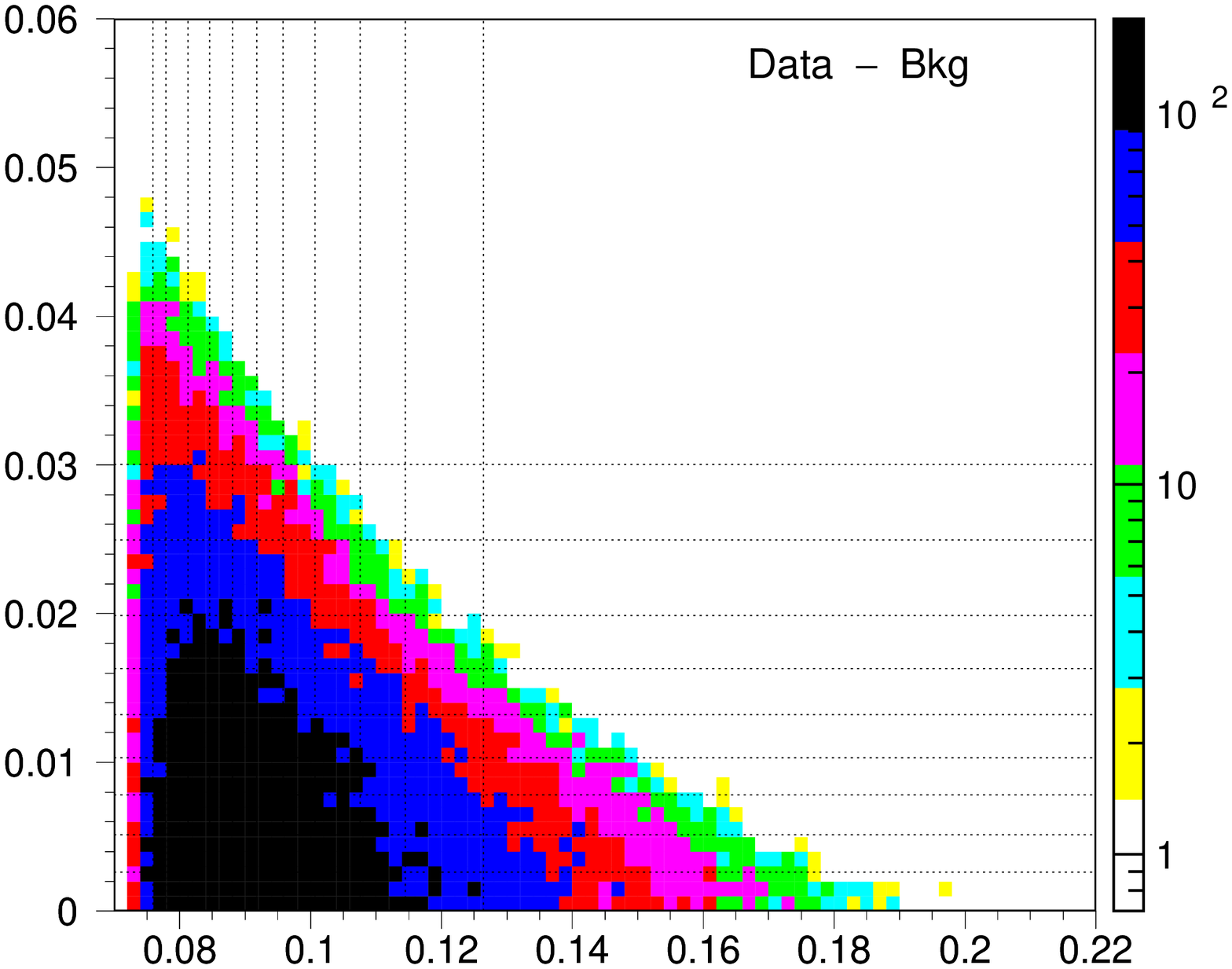}}
\put(10,-50){$\SP ~(\GEVcc)^2$}
\put(-135,150){$\SE ~(\GEVcc)^2$}
\put(40,130){\bf{ \large (a)}}
\put(90,-40){\epsfxsize78mm\epsfbox{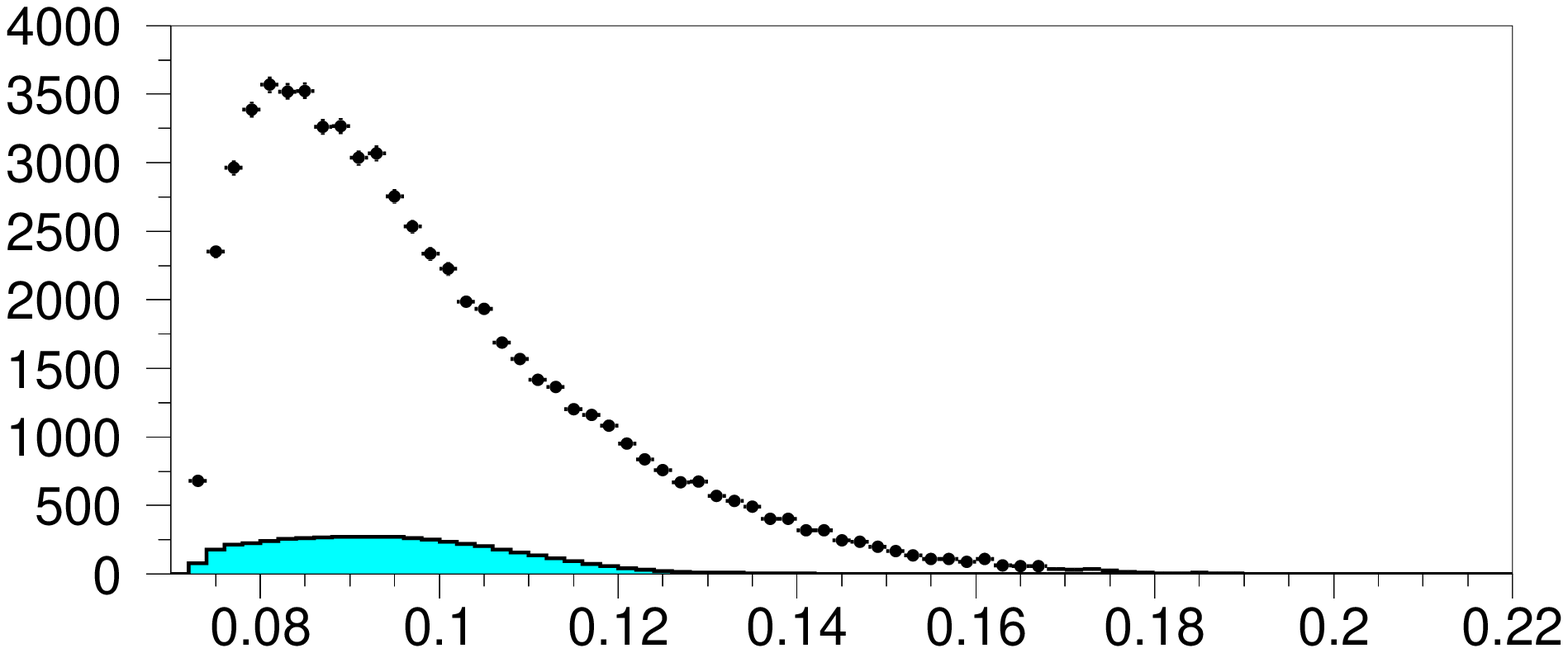}}
\put(240,50){$\SP ~(\GEVcc)^2$}
\put(272,130){\bf{ \large (b)}}
\put(120,-5){\epsfxsize78mm\epsfbox{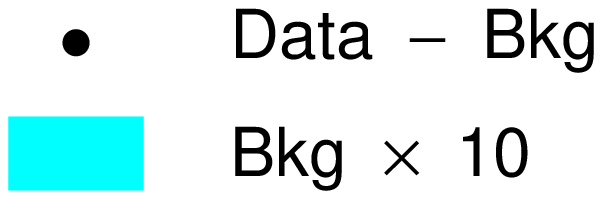}}
\put(90,-138){\epsfxsize78mm\epsfbox{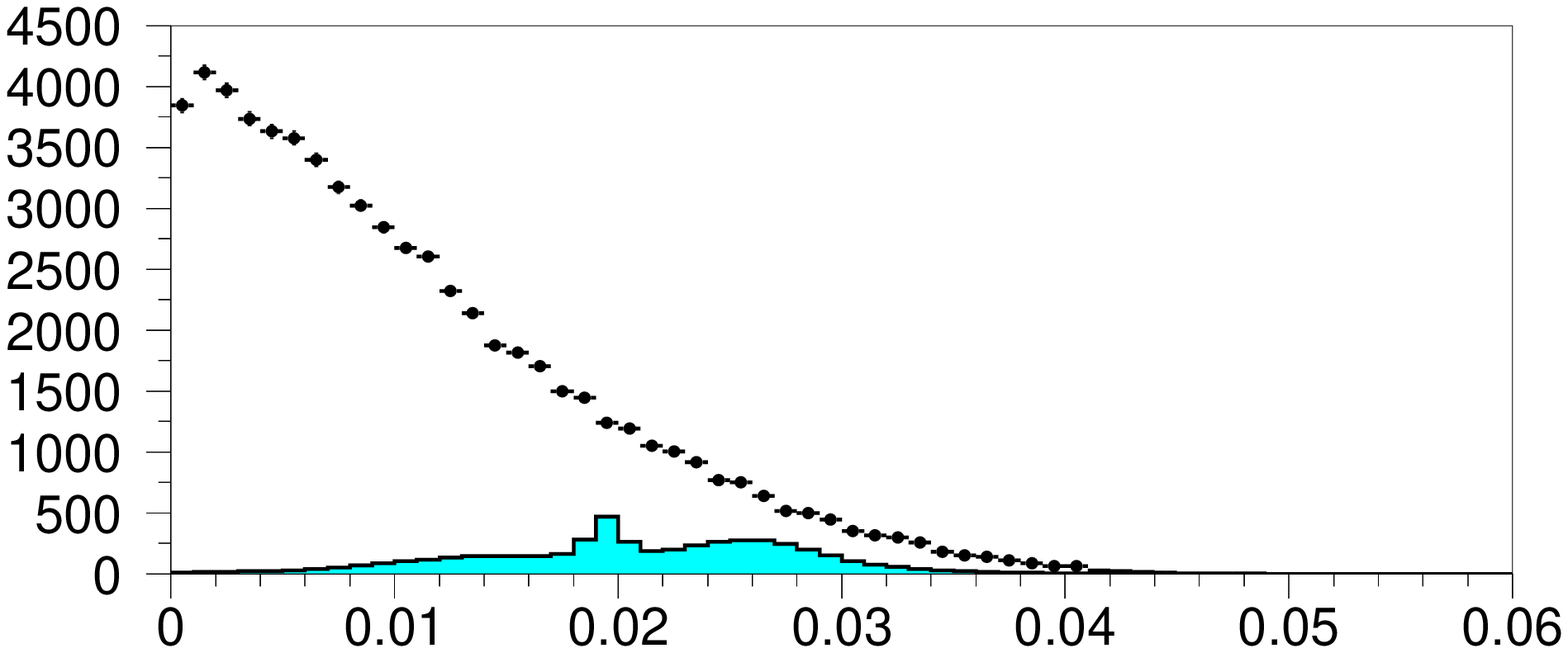}}
\put(240,-50){$\SE ~(\GEVcc)^2$}
\put(272,30){\bf{ \large (c)}}
\put(120,-105){\epsfxsize78mm\epsfbox{capfig8.eps}}
\end{picture}
\end{center}
\vspace{5mm}
\caption{{\bf (a)} Distribution of $\KEQO$ candidates in the plane 
($\SP,\SE$). The dotted lines represent the binning used in the fit procedure.
Distribution of $\KEQO$ candidates in the $\SP$ {\bf (b)} and $\SE$ {\bf (c)} 
variables after background subtraction (dots). Background, 
multiplied by a factor of ten to be visible, is displayed as a histogram.
\label{fig:spise}}
\end{figure}

Given the size of the data sample, a grid is defined with ten equal population
bins ($\sim 5900$ candidates per box) in the interval $\SP > 4 m_{\pi^+}^2$
and two equal population bins ($\sim 2900$ candidates per box) in the interval $\SP < 4 m_{\pi^+}^2$.
Along the $\SE$ variable, ten  bins of unequal width, common to all $\SP$ 
bins, are defined.
Eight boxes outside the kinematic boundary  are not populated and excluded from
the fit (Fig. \ref{fig:spise}a).

Dimensionless variables may also be used to describe the Dalitz plot, dividing
$\SP$ and $\SE$ by  $4 m_{\pi}^2$ with the arbitrary choice of using either
$m_{\pi^+}$ or $m_{\pi^{0}}$.
A natural choice would be to use $m_{\pi^{0}}$. For a direct comparison with 
the $\KEQC$ mode, it is however more appropriate to choose $m_{\pi^+}$.
The following variables are defined:
$q^2 = (\SP / 4 m_{\pi^{+}}^2 - 1) {\rm ~and~} 
y^2 = \SE/4 m_{\pi^{+}}^2 .$
The allowed kinematic range of $q^2$ for the $\KEQO$ decay spans both positive 
and negative values.
In a first approach, without any prior knowledge of the energy dependence,
an empirical parameterization (often called ``model independent'') is used
to describe the ratio of the data and simulated Dalitz plots. The ratio of 
the two distributions is equal to unity 
when the total number of simulated events, weighted for the 
best values of fit parameters, is normalized to the
total number of data events after background subtraction. This ratio
corresponds to $(F_s / f_s)^2$  where $f_s$ is a constant which can be
determined using the branching ratio measurement (see Section \ref{sec:ffabs}).
The fit procedure minimizes a $\chi^2$ expression in the two-dimensional space:
\begin{displaymath}
\chi^2 =  \sum_{i=1}^{12} \sum_{j=1}^{10} ((n_{ij} / m_{ij} - {\cal F}(q^2 _i ,y^2 _j,\hat{p})) / \sigma_{ij} )^2 ,
\end{displaymath}
where $n_{ij}$ is the number of background subtracted data events in the box 
$ij$, $m_{ij}$ is the number of simulated events observed in the same box for
a constant form factor, $\hat{p}$ is the set of fit parameters
and $ \sigma_{ij}$ is the
statistical uncertainty on the ratio, taking both data and simulation 
statistics into account.
The sum runs over the 112  boxes in the $(q^2 , y^2)$ space (12 bins along 
$q^2$ and 10 along $y^2$, excluding the 8 non-populated boxes).
The fit function ${\cal F}(q^2 , y^2 ,{\hat p})$ is defined as:
\begin{equation}\label{eq:fxy}
{\cal F}(q^2 , y^2 ,{\hat p}) = 
\begin{cases}
N (1 + a ~q^2 + b ~q^4 + c ~y^2 )^2 & {\rm ~for~} q^2 \geq 0 \\
N (1 + d ~C(q^2 ) + c ~y^2 )^2   & {\rm ~for~} q^2 < 0    
\end{cases}
~,
\end{equation}
where the fit parameters ${\hat p}$ are $(a,b,c,d$). 
The cusp-like function $C(q^2 )$ is defined as 
$C(q^2 ) = \sqrt{| q^2 / (1+q^2 )|}$ and $N$ is a normalization parameter. 
At each step of the fit, the function ${\cal F}$ is evaluated in each box at 
the corresponding reconstructed barycenter position $(q^2 _i , y^2 _i )$ using 
the 'true' $(q^2 , y^2 )$ values.  
The results from the 2-dimensional fit are given in Table \ref{tab:fsfe}.
The correlation matrix is symmetric and its non-diagonal terms are  
quoted in Table \ref{tab:correl}.

If the $\SE$ dependence is neglected ($c=0$), the fit quality becomes worse 
($\chi^2 / ndf = 129.8/108$, with a 7\% probability). The results of such a fit
are different from those obtained when considering a dependence on
$\SE$ and the degraded $\chi^2$ value supports the inclusion of an 
additional fit parameter. Other fits considering only data above $q^2 = 0$ have
also been performed with consistent results.
Omitting the $\SE$ dependence below $q^2 = 0$ brings a small increase of the 
$\chi^2$ value ($\chi^2 /ndf$ becomes 107.9/107 with a reduced $46\%$
probability compared to Table \ref{tab:fsfe}). Allowing the $\SE$
dependence to be different above and below $q^2 = 0$ leads to very close 
values with larger errors and no $\chi^2$ improvement. Therefore the  
formulation of Eq. (\ref{eq:fxy}) with an identical $y^2$ dependence below and 
above $q^2 = 0$ is considered in the final result.

\begin{table}[htp]
\caption{\label{tab:fsfe} 
Result of the fit to the Dalitz plot. The coefficients are defined in
 Eq. (\ref{eq:fxy}). The errors are statistical only.}
\begin{center} 
\vspace{-10mm}
\begin{tabular}{l|  r }
 fit parameters & \multicolumn{1}{c}{values} \\
\hline
$a$ & $ 0.149 \pm 0.033$ \\
$b$ & $-0.070 \pm 0.039$ \\
$c$ & $ 0.113 \pm 0.022$ \\
$d$ & $-0.256 \pm 0.049$ \\
$N$ & $0.0342 \pm 0.0004$ \\
$\chi^2 / ndf$ & $101.4/107 = 0.95$ \\
probability & $63\%$ \\
\end{tabular}
\vspace{-16mm}
\end{center}
\end{table}
\begin{table}[htp]
\caption{\label{tab:correl}
Non-diagonal correlation coefficients between the two-dimensional fit 
parameters.}
\begin{center} 
\vspace{-8mm}
\begin{tabular}{l|  r r r r}
        &     $a$  &  $b$     &   $c$     & $d$ \\
\hline
 $N$    & $-0.751$ & $0.581$  & $-0.644$ & $-0.417$ \\
 $a$    &          & $-0.946$ & $ 0.180$ & $0.467$ \\
 $b$    &          &          & $-0.062$ & $-0.400$ \\
 $c$    &          &          &          & $-0.028$  \\
\end{tabular}
\vspace{-10mm}
\end{center}
\end{table}

\begin{figure}[ht]
\begin{center} 
\vspace{4mm}
\begin{picture}(160,60)
\put(-140,-130){\epsfxsize80mm\epsfbox{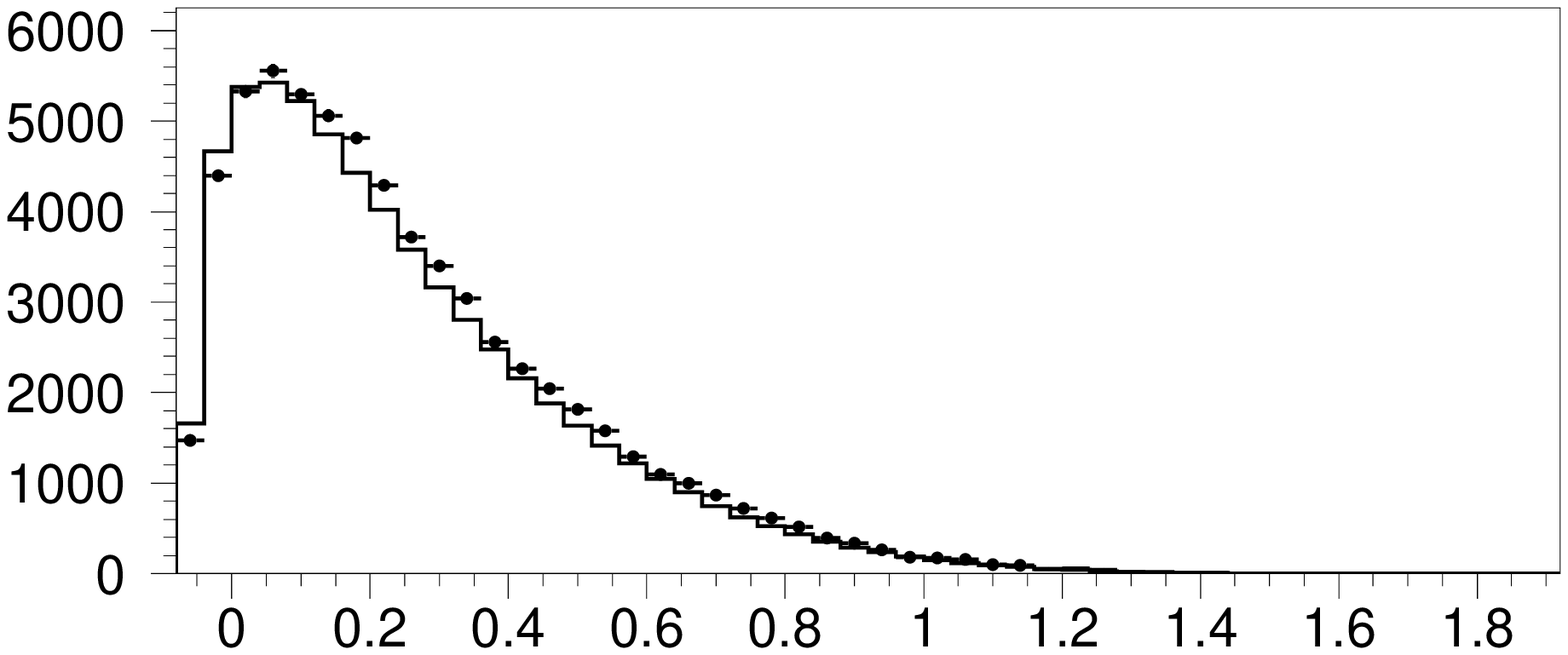}}
\put(-118,-85){\epsfxsize80mm\epsfbox{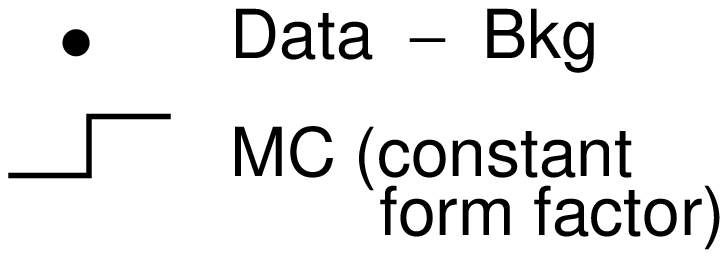}}
\put(80,-130){\epsfxsize80mm\epsfbox{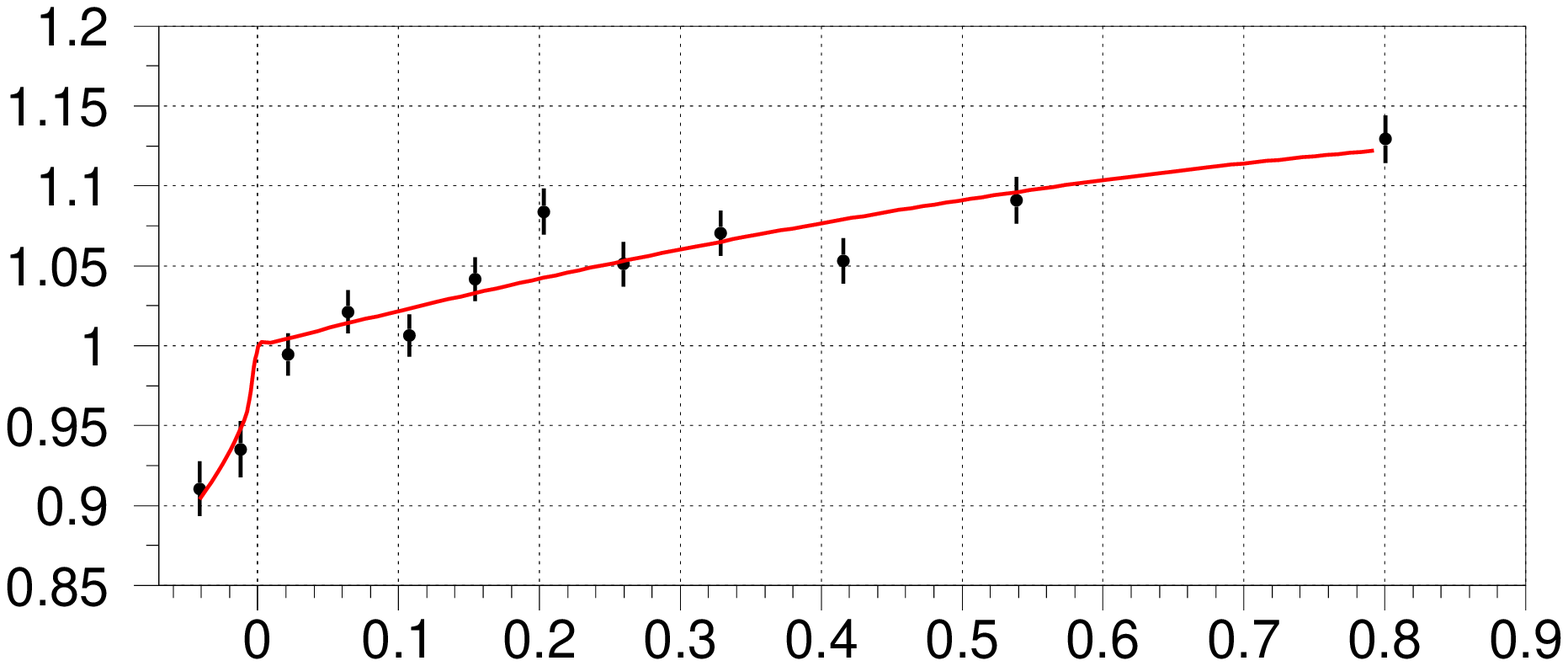}}
\put(90,67){$(F_s / f_s)^2 / N$}
\put(50,-40){$q^2$}
\put(47,45){\bf{\large (a)}}
\put(280,-40){$q^2$}
\put(110,45){\bf{\large (b)}}
\end{picture}
\end{center}
\caption{{\bf (a)} Data and simulation (including a constant form factor) 
$q^2$ distributions 
$(q^2 = \SP / 4 m_{\pi^{+}}^2 - 1)$.
{\bf (b)} Ratio of the two $q^2$ distributions in equal population bins.   
Each symbol is plotted at the barycenter position of the data events in the 
bin to account correctly for the variable size binning. 
The line corresponds to the empirical description using the best fit-parameter 
values:
a degree-2 polynomial above $q^2 = 0$ and a cusp-like function below.
\label{fig:ffq2}}
\end{figure}

For a simpler display, the projection of the Dalitz plot on
the $q^2$ variable is shown in Fig. \ref{fig:ffq2} for
data and simulation (generated with a constant form factor).
As the $q^2$ distribution is very steep at negative values, the comparison
is also shown as the ratio of the two distributions in equal population bins:
the statistical errors are
identical for the last 10 equal population bins and larger by a factor of 
$\sqrt{2}$ for the first two bins (half population).

\begin{figure}[htp]
\begin{center}
\begin{picture}(160,200)
\put(-100,-40){\epsfxsize120mm\epsfbox{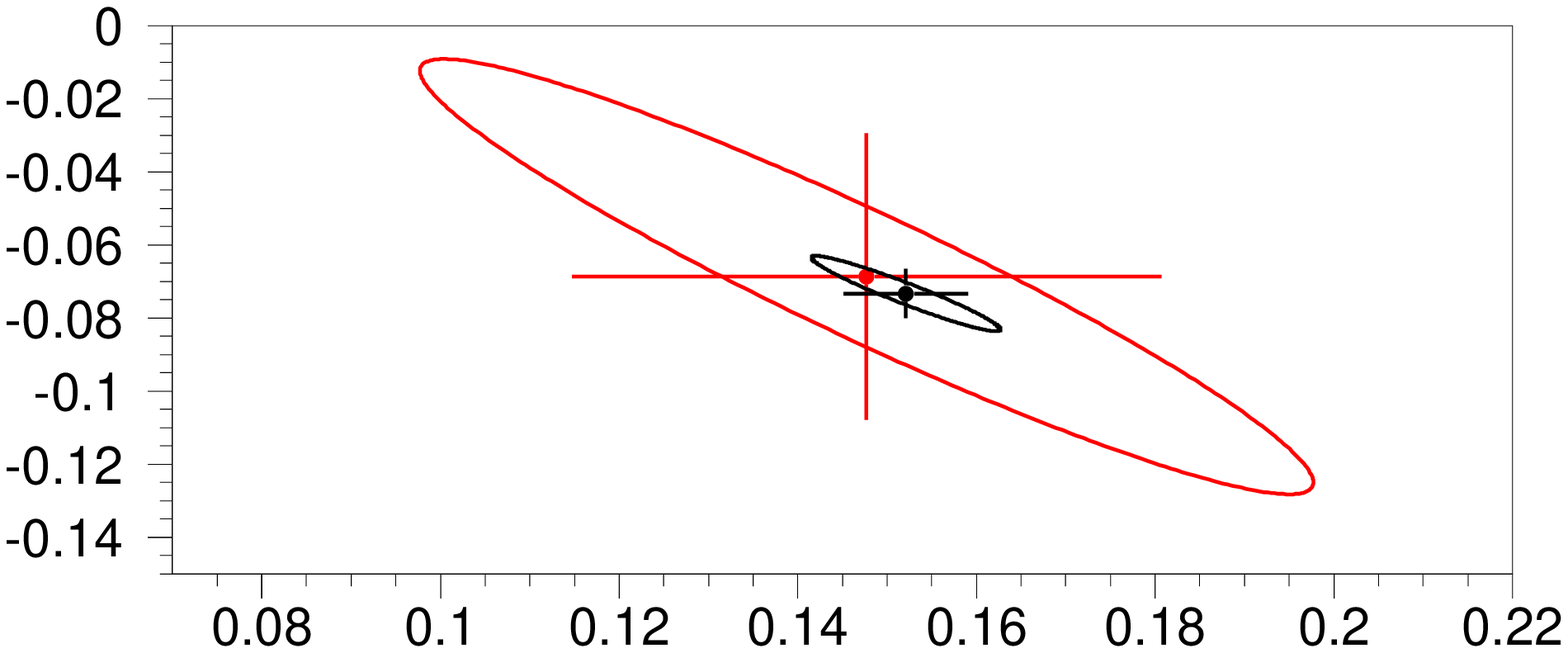}}
\put(-120,-180){\epsfxsize120mm\epsfbox{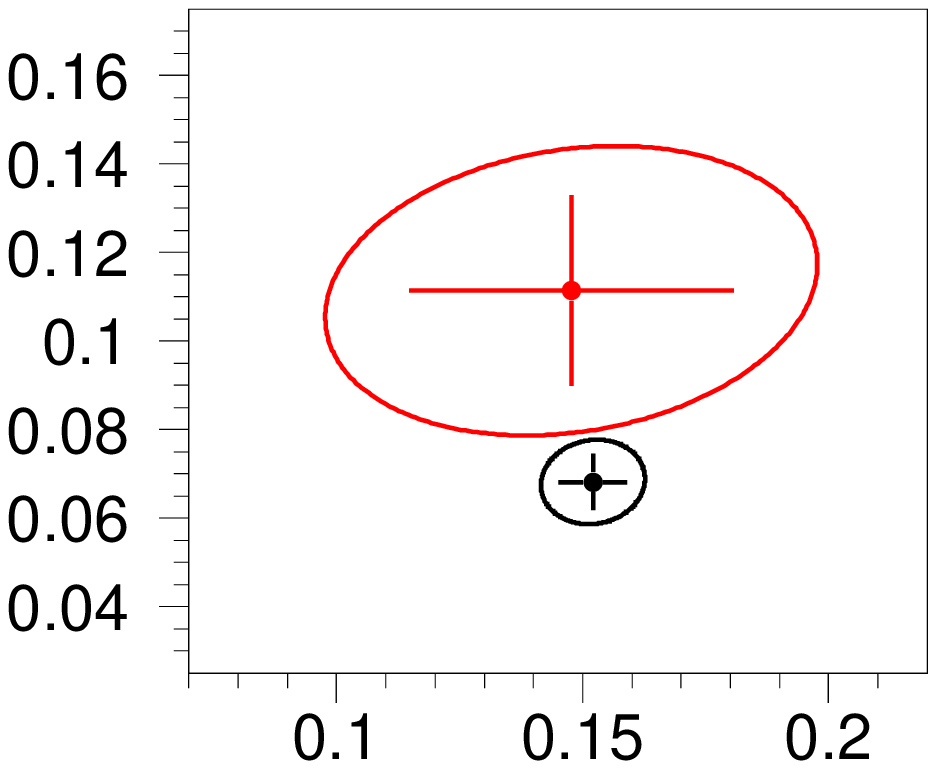}}
\put(75,-180){\epsfxsize120mm\epsfbox{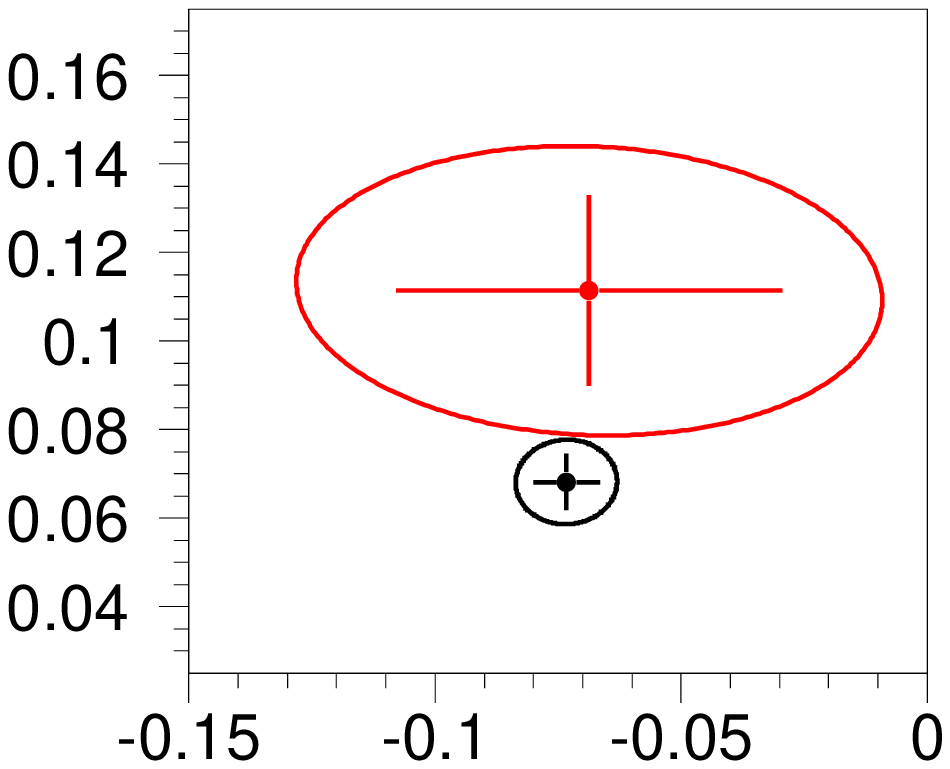}}
\put(-135,240){$b = f^{\prime\prime}_s / f_s$} 
\put(-135,100){$c = f^{\prime}_e / f_s$} 
\put(60,100){$c = f^{\prime}_e / f_s$} 
\put(250,118){$a = f^{\prime}_s / f_s$}  
\put(250,-23){$b = f^{\prime\prime}_s / f_s$}
\put(42,-23){$a = f^{\prime}_s / f_s$}
\end{picture}
\end{center}
\vspace{-5mm}
\caption{Form factor description in 2-parameter planes obtained in $\KEQO$ and
$\KEQC$ analyses in the $(q^2 , y^2)$ series expansion formulation. The top 
plot corresponds to the 
$(f^{\prime}_s / f_s , f^{\prime\prime}_s / f_s)$ plane (the $(a,b)$ 
plane), the bottom plots to the $(f^{\prime}_s / f_s ,f^{\prime}_e / f_s)$ and 
$(f^{\prime\prime}_s / f_s, f^{\prime}_e / f_s)$ planes (the $(a,c)$
and $(b,c)$ planes, respectively). Errors plotted are statistical only and all
contours are $68\%$ CL. The smaller area corresponds to the $\KEQC$
result obtained from a large statistical sample \cite{ke410}.
The correlations between fitted parameter errors are very similar and results 
are consistent within statistical errors.
\label{fig:ellipses}}
\end{figure}

The results in the $(q^2, y^2)$ formulation can be directly
compared to those obtained in the $\KEQC$ analysis \cite{ke410}  where the 
corresponding form factor is described as 
$F_s = f_s (1 + f^{\prime}_s / f_s ~q^2 + f^{\prime\prime}_s / f_s ~q^4 
+ f^{\prime}_e / f_s ~y^2 )$. They are displayed in Fig. \ref{fig:ellipses} in 
the three 2-parameter planes.
\vspace{18mm}
\clearpage
\subsection{Systematic uncertainties}
Many possible sources of systematics uncertainties have been explored and 
details are given for the main contributions. 

{\bf Background control}.
Background has been studied both in shape and rate across the Dalitz plot. 
The most sensitive item is the fake-electron background from $\KTPO$.

The shape of the background can be modified by extending further out the 
ellipse cut in the $(M_{3\pi}, p_t)$ plane (Section \ref{sec:selevt}): due to 
the location of the
fake-electron background close to the ellipse cut boundary, its fraction
varies rapidly from 0.65\% to 0.50\%, 0.39\% and 0.31\% when increasing the
ellipse main axes by 10\%, 20\% and 30\% of their nominal values while the 
signal loss (estimated from simulation) is 1.2\%, 2.7\% and and 4.4\%, 
respectively. This 
changes both the rate and the shape of the fake-electron background while the 
relative fraction of background from $\PIEN$ decays (0.12\%) and
accidentals (0.22\%) are unaffected. 

The use of looser or tighter electron-identification criteria is another 
way to vary the background contribution both in shape and relative rate. The 
efficiency of the DV cut as a function of the cut value is well known from 
previous studies \cite{ke410,ke412}. 
When changing the cut value from 0.90 (reference) to 0.85 (0.95), the number
of candidates changes by $+1.1\% ~(-2.8\%)$, the fraction of 
fake-electrons relative to signal changes from 0.65\% to 0.78\% (0.47\%) 
while the relative fraction of 
decay-electrons remains unchanged (0.12\%). Conservatively, the maximum 
difference observed between any of the five fit results and the reference value
is quoted as a systematic uncertainty, not taking into account the large 
anti-correlation between parameters $a$ and $b$ (Table \ref{tab:correl}). The 
quoted contribution is then ${\cal O} (1 \times 10^{-2})$ for all parameters.

The extrapolation method (Section \ref{sec:bkg}) using a single scaling 
factor or a momentum dependent factor has little impact (few $10^{-4}$) as the 
momentum spectra of control regions C and D are very similar. 

The fit parameters vary linearly with the rate of each background component as
observed when scaling each nominal component by a factor of 0, 0.5, 1, 2 while
keeping its shape unchanged. 
The fake-electron background ($425 \pm 2$ events) is known to better than 1\%,
the decay-electron component ($79 \pm 1$ events) is known to about 1\% and the 
accidental background ($146 \pm 13$ events) to about 10\%. The uncertainty 
related to each background scale is a few $10^{-4}$ (or less) for all 
parameters.

All considered contributions are then added in quadrature and the sum quoted in
Table \ref{tab:ffsumsys}.

{\bf Radiative events modeling}.
The $\KEQO$ final state contains at least four photons. 
To evaluate how the presence of additional photons can distort the
measurement (either from Inner Bremsstrahlung (IB) at the decay vertex or from 
External Bremsstrahlung (EB) emitted in the interaction of the $e^{\pm}$
with matter), dedicated simulated samples without IB (or EB) photon 
emission have been analyzed as real data. 

Other studies \cite{ebmatter} have shown that the material description before 
the spectrometer magnet in terms of radiation length
is known within 1.1\% precision. One percent of the full effect 
observed when omitting EB is quoted as 
a systematic error of few $10^{-4}$ for all parameters.

As reported by the {\tt PHOTOS} authors in Ref. \cite{ke4rad}, the IB modeling 
uncertainty should not exceed 10\% of the full effect. Therefore 10\% of 
the difference between the results obtained with and without IB is quoted as 
an uncertainty on the photon emission modeling with a few ${\cal O} (1 \times 
10^{-3})$ contribution for all fit parameters.
Both EB and IB contributions are added in quadrature, dominated by the IB
modeling uncertainty.
 
{\bf Others}.
The analysis of simulated samples, different from those used in the fit and 
treated as real data, has not revealed any bias in the fit procedure.  
The variation of the chosen grid in $\SE$ has no significant impact on the fit 
results.

Applying more stringent criteria in the reconstruction, excluding either
5\% of candidates having more than one vertex solution, 
or 1.2\% of candidates with no available track HOD time, 
or up to 20\% of candidates with a reconstructed $\MEN$ value lower than 60
$\MEVcc$ (affected by a worse resolution), 
shows no significant effect on the fit results.

When the corrections applied to the simulation samples are removed in turn, the
only sizable change is observed when omitting the variation of the MBX 
trigger across the Dalitz plot. The studies of abundant $\KTPO$ events have 
shown a good agreement between data and simulation within 1\%. One percent of
the difference between the results obtained with and without MBX trigger 
simulation is quoted as systematic uncertainty.
 
The offline $M_{miss}$ cut is chosen to be more strict than the online 
trigger requirement to guarantee high efficiency. Moving further
away from the nominal cut (206 $\MEVcc$, see Section \ref{sec:selevt}) to 217, 
227 and 237 $\MEVcc$, the signal statistics decreases by 5.4\%, 11\% and 17\%, 
respectively. The fit results are in agreement with the nominal analysis 
results within the statistical errors and with no definite trend. 

{\bf Acceptance control stability}.
Stability checks are performed by
splitting the data sample into statistically independent subsamples and
comparing fit results. Two independent subsamples are defined according to each
quantity to keep statistical errors low enough and the split is repeated for 
many different quantities.
These studies investigate possible biases from lack of control of the beam
geometry (achromat polarity, kaon beam charge), the spectrometer and
calorimeter response (spectrometer magnet polarity, regions of LKr 
geometrical illumination from track transverse position), the detector geometry
(vertex $Z$ position) and their overall time variation (year).
The 14 fits are obtained with a good quality $\chi^2$ and the parameter 
variations are consistent within the increased statistical error and the 
correlation matrix. 

A summary of all contributions from studied effects is given in
Table~\ref{tab:ffsumsys}. The main contribution comes from the background
control while the radiative events modeling contribution is much smaller. Other
sources give marginal contributions.
\begin{table}[htp]
\caption{\label{tab:ffsumsys} Systematic uncertainty contributions to the form
factor description. The parameter values and their statistical error are also 
recalled for completeness.}
\begin{center}
\vspace{-10mm}
\begin{tabular}{l| c c c c}
Source  & $\delta a$ & $\delta b$ &  $\delta c$ & $\delta d$ \\
\hline
 Background control        & 0.0140  & 0.0122  & 0.0062  & 0.0164\\
 Radiative events modeling & 0.0037  & 0.0035  & 0.0033  & 0.0013 \\
 Fit procedure  &  -- &  -- &  -- &  --\\
 Reconstruction/resolution & -- &  -- &  -- &  -- \\
 Trigger simulation        &$< 0.0001$ & $< 0.0001$ & $< 0.0001$ & $< 0.0001$\\
 Acceptance control        &  -- &  -- &  -- &  -- \\
\hline
 Total systematics & \phantom{0}0.014 & \phantom{-}0.013 & \phantom{0}0.007 & \phantom{-}0.016\\
\hline
\hline
Parameter   & $a$   & $b$       &  $c$  & $d$ \\
 Value      & 0.149 &  $-0.070$ & 0.113 & $-0.256$ \\
 Statistical error & 0.033 & $\phantom{-}0.039$ & 0.022 & $\phantom{-}0.049$
\end{tabular}
\vspace{-15mm}
\end{center}
\end{table}

\subsection{Discussion}
\begin{figure}[htp]
\begin{center}
\begin{picture}(160,150)
\put(-185,-85){\epsfxsize100mm\epsfbox{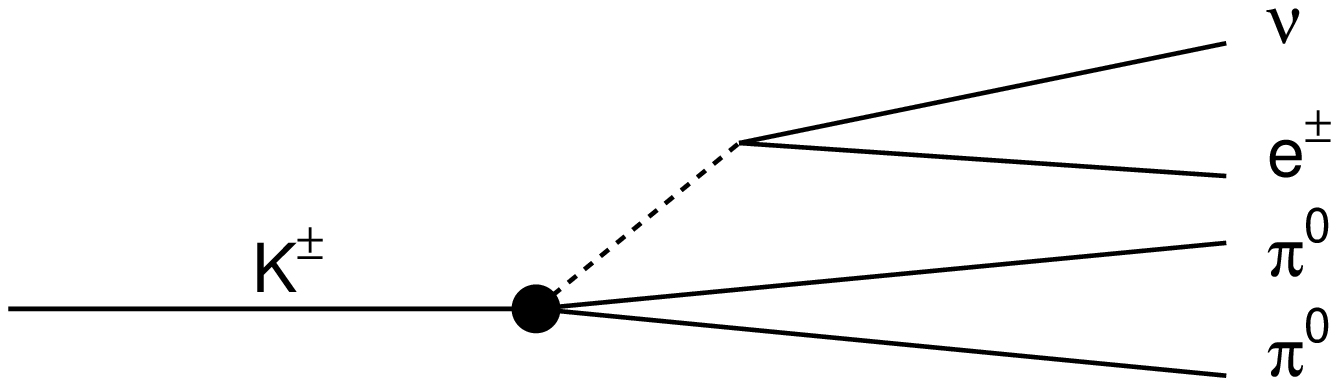}}
\put(70,-85){\epsfxsize100mm\epsfbox{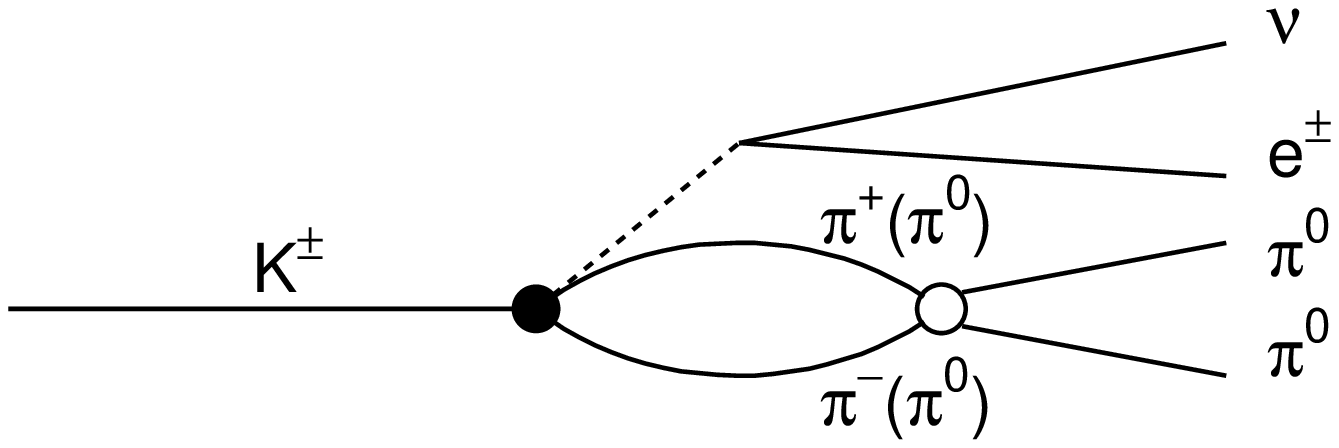}}
\put(-145,130){\bf \large (a)}
\put(110,130){\bf \large (b)}
\end{picture}
\end{center}
\vspace{-4cm}
\caption{{\bf (a)} Tree level diagram (amplitude ${\cal M}_0$) for the $\KEQO$
 decay mode.
{\bf (b)} One-loop diagram (amplitude ${\cal M}_1$) with contribution of  the 
$\KEQC$ decay mode to the $\KEQO$ final state.
\label{fig:diag}}
\end{figure}

The observed deficit of events at  $q^2 < 0$
can be related to the final state charge exchange scattering process 
$(\PPI \MPI \ra \PIo \PIo)$ in the $\KEQC$ decay mode.
In a naive and qualitative approach, one may take advantage of the early 
one-loop description of re-scattering effects in the $\KTPO$ mode
\cite{cabibbo} and consider a similar interpretation in the $\KEQO$ decay mode,
defining the tree level amplitude ${\cal M}_0$ 
and the one-loop amplitude ${\cal M}_1$ (Fig.~\ref{fig:diag}) of the $\KEQO$ mode.

The tree level amplitude ${\cal M}_0$ has a dispersive behavior above and
below $q^2 = 0$. The one-loop amplitude ${\cal M}_1$ is imaginary for $q^2 > 0$
$(i{\cal M}_1)$ and real for $q^2 < 0$ $({\cal M}_1)$. It has two components:
a dispersive component which can be absorbed in the unperturbed 
amplitude ${\cal M}_0$ and a negative  absorptive component.
The total amplitude squared is then written as:
\[
 |{\cal M}|^2 =     
\begin{cases}
|{\cal M}_0 + i {\cal M}_1|^2 = ({\cal M}_0 ) ^2 + ({\cal M}_1) ^2 &  q^2 > 0 \\
|{\cal M}_0 + ~{\cal M}_1 |^2 = ({\cal M}_0 )^2 + ({\cal M}_1 )^2 + 2 {\cal M}_0 {\cal M}_1 ,& q^2 < 0
\end{cases}
 ~.
\]

In this approximate approach, neglecting a potential $\SE$ dependence, 
${\cal M}_0$ can
be developed in a series expansion in $q^2$ (as described for $q^2 >0$ in
the form factor measurement in Section \ref{sec:ff} and Table \ref{tab:fsfe})
and  ${\cal M}_1$  can be expressed as:
\begin{displaymath}
{\cal M}_1 =  - 2/3 ~(a_0 ^0 - a_0 ^2) ~F^{+-}_s ~\sigma_{\pi}(q^2 ),
\end{displaymath}
where $F^{+-}_s = f^{+-}_s (1 + f^{\prime}_s / f_s ~q^2 + f^{\prime\prime}_s / f_s ~q^4)$
is the $\KEQC$ form factor \cite{ke410}, $a_0 ^0$  and $a_0 ^2$ 
are the S-wave $\pi \pi$ scattering lengths in the isospin states $I=0$ and
$I=2$, while
$\sigma_{\pi}(q^2 ) = \sqrt{1 - 4m _{\pi^+}^{2}/\SP} =\sqrt{| q^2 /(1+q^2 )|}$
introduces, through the interference term below $q^2 = 0$,
a cusp-like behavior as observed in the data.

Better descriptions of re-scattering effects in the $\KTPO$ decay amplitude 
already exist, including two-loop effects \cite{cabisi} and also radiative
corrections within a ChPT calculation \cite{theocusp}. Recent developments
on the related topic of the low energy pion form factors \cite{knecht} may also
bring a more elaborate description of the $\KEQO$ amplitude including two-loop 
contributions, $\pi\pi$ scattering and mass related isospin symmetry breaking 
effects. Once available, such an approach could be exploited further to extract
more information related to physical quantities from the result reported here.

\section{Branching ratio measurement}
\subsection{Inputs}
All input ingredients to the BR($\KEQO$) measurement (Eq. (\ref{eq:br})) are 
summarized in Table \ref{tab:br1} for each kaon charge, summed over the ten 
subsamples (or averaged when appropriate) while the final result is obtained as
the statistical average of the ten independent subsamples summed over both kaon
charges (Fig. \ref{fig:brperss}).
Because of the symmetrization of the beam and detector geometries, the
global $\KPL$ and $\KMI$ acceptances are very similar: $\KPL$ and $\KMI$ beam 
lines are exchanged when inverting the achromat polarity while positive and 
negative charged track trajectories follow similar paths in the spectrometer 
when inverting the spectrometer magnet polarity. Data taking conditions have 
been set up carefully to equalize the integrated kaon flux in the four 
configurations of achromat and spectrometer magnet polarities \cite{agcomb}.
The acceptances $A_{s}$ and $A_{n}$ are obtained using the most elaborate
description of the decay dynamics, in particular the model independent
parameterization of the signal form factor reported here.
The trigger efficiencies $\vep_{n}$ ($\vep_{s}$) are the product of the two
measured trigger components NUT and Q1$\cdot$MBX. All quoted uncertainties are 
of statistical origin.
\begin{table}[htp]
\caption{\label{tab:br1} Inputs to the BR($\KEQO$) measurement for each kaon
charge summed over subsamples. Uncertainties on the last digits are given
in parentheses. The last two columns display the overall numbers and the
contribution of each component to the relative branching ratio error.}
\begin{center}
\vspace{-6mm}
\begin{tabular}{l| r r || r   c }
   &   $\KPL$   &   $\KMI$  &  $\KPM$  & $\delta$BR/BR $(\times 10^{4})$ \\
\hline
 $N_{s}$    &     41 850   &     23 360   &     65 210 & 39 \\
 $N_{b}(s)$ &        418   &        233   &        651 &  4 \\
 $N_{n}$    & 60 107 311   & 33 436 659   & 93 543 970 &  1 \\
 $N_{b}(n)$ &    148 486   &     82 600   &    231 086 &  $\ll 1$ \\
 $A_{s}$    & $1.927(2)\%$ & $1.923(2)\%$ & $1.926(1)\%$ & 5 \\
 $A_{n}$    & $4.053(2)\%$ & $4.047(3)\%$   &  $4.052(2)\%$ & 5 \\
\hline
 $\vep_{s}$ &\multicolumn{3}{c}{$96.06(3)\%$}  & 3 \\
 $\vep_{n}$ &\multicolumn{3}{c}{$97.42(0)\%$}  & $\ll 1$ \\
\hline
\multicolumn{4}{l}{Total relative error}      & 40 
\end{tabular}
\end{center}
\vspace{-10mm}
\end{table}
\subsection{Systematic uncertainties}
Some sources of uncertainty are expected to affect the 
corresponding quantities for signal and normalization modes in a similar way 
and therefore have a limited impact as they cancel at first order in the ratio
of Eq. (\ref{eq:br}). Some others are specific to the signal or normalization 
mode.

{\bf Background in the \boldmath $\KEQO$ \unboldmath sample.}
The fake-electron component ($425 \pm 2$ events) is obtained with an 
uncertainty from the extrapolation procedure of 0.4\%.
Conservatively, the  half difference between the evaluations 
based on two control subregions (restricted to $E/p$ ranges from 0.2 to 0.45 
and from 0.45 to 0.7, respectively, see Fig. \ref{fig:fak}) 
is assigned as an additional systematic error of $\pm 5$ events and added in
quadrature. This background contributes $\delta$BR/BR $= 1 \times 10^{-4}$.

The uncertainty on the $\PIEN$ component ($79 \pm 0.7$ events) is due to the 
limited statistics of the simulation and BR$(\PIEN)$ precision, adding up to 
$0.8\% \oplus 0.3\% = 0.9\%$. This contribution $\delta$BR/BR 
$= 0.1 \times 10^{-4}$ is marginal. 

The accidental component precision ($146 \pm 12$ events) is limited by the 
statistics of the side band signal sample. The statistical error is quoted
as systematics and contributes $\delta$BR/BR $=2 \times 10^{-4}$.
Adding in quadrature the three contributions, the background systematic 
uncertainty is $\delta$BR/BR = $2.2 \times 10^{-4}$.

An estimation of the uncertainty in the electron identification procedure is 
obtained from the stability of the result when varying the DV cut value between
0.85 and 0.95, changing the fake-electron background by a factor close to 2 
(Fig. \ref{fig:lda}b). 
The analysis of the signal mode was repeated for three cut values 
(0.85, 0.90, 0.95) and the observed change quoted as $\delta$BR/BR = 
$25 \times 10^{-4}$, the dominant contribution.

{\bf Radiative effects.}
The event selection requires a minimum track to photon and 
photon to photon distance at the LKr front face. The precise description
of IB (EB) emission may affect the acceptance calculation.

Dedicated MC samples simulated without IB photon emission 
are used to estimate the impact of the {\tt PHOTOS} description.
The signal acceptance $A_s$ increases by $1.9\%$ while $A_n$ is unchanged. 
One tenth of the observed effect is assigned as a modeling uncertainty 
according to the prescription of Ref. \cite{ke4rad}, $\delta$BR/BR = 
$19\times 10^{-4}$.

Dedicated simulations including IB by {\tt PHOTOS}, 
but switching off EB in the GEANT tracking in the detector, show that $A_n$ is
unaffected (within the simulated statistics) while $A_s$ increases by 
$(3.5 \pm 0.4)\%$.
The agreement between data and simulation in term of radiation length 
is quoted as 1\% as studied in Ref. \cite{ebmatter}. 
This fraction of the observed change is propagated as 
$\delta$BR/BR = $4\times 10^{-4}$
and is added in quadrature to the dominant IB-related uncertainty.  

{\bf Form factor description in the \boldmath $\KEQO$ \unboldmath simulation.}
The signal acceptance $A_s$ calculation depends on the form factor description 
considered in the simulation.
When using $F_s$ descriptions from NA48/2 $\KEQC$ \cite{ke410} or $\KEQO$ 
(present work) modes,
$A_s$ changes by less than 0.01\% 
consistent with no change within the corresponding statistical precision.
Both descriptions are in agreement while the $\KEQC$ form factor coefficients
are obtained with better precision (Fig. \ref{fig:ellipses}).
Moving each coefficient in turn by $\pm 1 \sigma$ away from its measured 
value, the corresponding acceptance variations are obtained. 
Conservatively (i.e. neglecting the anti-correlation between the $a,b$ fit 
parameters), these $A_s$ variations related to the $a,b,c$ coefficients in
the $\KEQC$ mode \cite{ke410} are added in quadrature.
The major contributions come from $a$ and $b$ parameter variations.
The acceptance has about three times less sensitivity to the $c$ coefficient
and about twelve times less to the $d$ coefficient. Therefore
including or not the $d$ contribution 
does not change the quoted uncertainty  
$\delta$BR/BR = $17 \times 10^{-4}$. 

{\bf Acceptance stability.}
Many stability checks have been performed varying the selection cuts. 
The acceptances $A_s$ and $A_n$ are particularly sensitive to the minimum 
radial track position at DCH1 ($R_{CH1}$).
When increasing $R_{CH1}$ by steps of 1 cm, the number of 
$\KEQO$ candidates decreases by steps of about 3\% and the number of $\KTPO$ 
candidates by larger steps of about 4\%. 
Changes in acceptance and number of candidates largely compensate 
each other in the BR calculation. Therefore only the largest 
significant difference is quoted as the corresponding uncertainty, 
$\delta$BR/BR = $16 \times 10^{-4}$.

The control level of the time variation of the acceptance is estimated by 
swapping the acceptances (obtained from simulation) of pairs of subsamples 
recorded during different time periods. This 
leads to a conservative estimate $\delta$BR/BR = $4 \times 10^{-4}$. 
 
The stability of the BR value with the spectrometer magnet polarity 
(B$^{+}$, B$^{-}$), the achromat polarity (A$^{+}$, A$^{-}$), the year of data 
taking (2003, 2004) or the kaon charge (K$^{+}$, K$^{-}$)
has not revealed any significant effect. 

The above effects are combined into 
$\delta$BR/BR  = $16 \times 10^{-4}$.

{\bf Level 2 trigger cut.}
Varying the $M_{miss}$ cut applied in the selection and recomputing both 
acceptances and trigger efficiencies provide an estimate of the
uncertainty related to the trigger cut. Moving the cut value from 206 to 
227 $\MEVcc$ in the selection, the acceptance, trigger efficiency and number 
of candidates in the normalization sample are unaffected.
The signal statistics decreases by 12.5\% at no gain in the trigger 
efficiency and therefore will only increase the statistical error by 6\%.
The difference between the branching ratio values obtained for both cut values
is quoted as a systematic uncertainty $\delta$BR/BR  = $4 \times 10^{-4}$.

{\bf Beam geometry modeling and resolution.}
The comparison of the reconstructed parent kaon momentum distributions of data 
and simulated $\KTPO$ candidates can be used to improve the beam geometry 
modeling. 
This fine tuning of the beam properties is propagated to the $\KEQO$
simulation.  
As the selection cuts are loose enough, there is little sensitivity to these 
mismatches and the observed change of $A_s$ is negligible.

{\bf Spectrometer and calorimeter calibrations.}
The study of the mean reconstructed $\KTPO$ mass as a function of the charged 
pion momentum and of the photon energies is an indication of the level of 
control of the spectrometer momentum calibration and the calorimeter energy 
calibration.

Both data and simulated reconstructed $M_{3\pi}$  distributions 
show a similar residual variation with the charged pion momentum,  
which indicates that the momentum calibration could still 
be improved. 
However, the maximum effect of $\pm 0.35 ~\MEVcc$ is well below the achieved 
resolution of $1.4~\MEVcc$. 
The residual variations with the photon energies are also similar for data and 
simulated samples and within $\pm 0.35 ~\MEVcc$,
consistent with a relative 
change in energy scale smaller than $1\times 10^{-3}$.
No additional systematic uncertainty is assigned.
 
{\bf Simulation statistics and trigger efficiency.} 
Acceptances and trigger efficiencies are already quoted in Table \ref{tab:br1}.
Their statistical errors are propagated as systematic errors.
Errors on $A_s$ and $A_n$ are due to the limited size of the simulation samples
and added in quadrature.
The combined error from $\vep_{s}$ and $\vep_{n}$ is dominated by the precision
on $\vep_{s}$.

Table~\ref{tab:brsys} summarizes the considered contributions. The external
error comes from the uncertainty on BR($\KTPO$) in the normalization mode.

\begin{table}[htp]
\vspace{-3mm}
\begin{center}
\caption{Summary of the relative contributions to the BR$(\KEQO)$ systematic
uncertainty. For completeness, uncertainties related to simulation statistics 
and trigger efficiencies are also quoted here globally while they are treated 
in the analysis as time dependent errors of statistical origin.
\label{tab:brsys}}
\vspace{2mm}
\begin{tabular}{lcc}
Source              & $\delta$BR/BR $\times 10^{2}$ \\
\hline
Background and electron-ID               & 0.25 \\
Radiative events modeling & 0.19 \\
Form factor uncertainty   & 0.17 \\
Acceptance stability      & 0.16 \\
Level 2 Trigger cut       & 0.04 \\
Simulation statistics     & 0.07 \\
Trigger efficiency        & 0.03 \\
\hline
Total  systematics        & 0.40 \\
\hline
External error from BR($\KTPO$) & 1.25 \\
Statistical error         & 0.39 \\
\end{tabular}
\vspace{-1cm}
\end{center}
\end{table}

\subsection{Results}
\begin{figure}[htp]
\begin{center}
\begin{picture}(160,200)
\put(-125,180){ \large  BR($\KEQO) \times 10^5$ }
\put(190,30){ sample number}
\put(-120,-120){\epsfxsize120mm\epsfbox{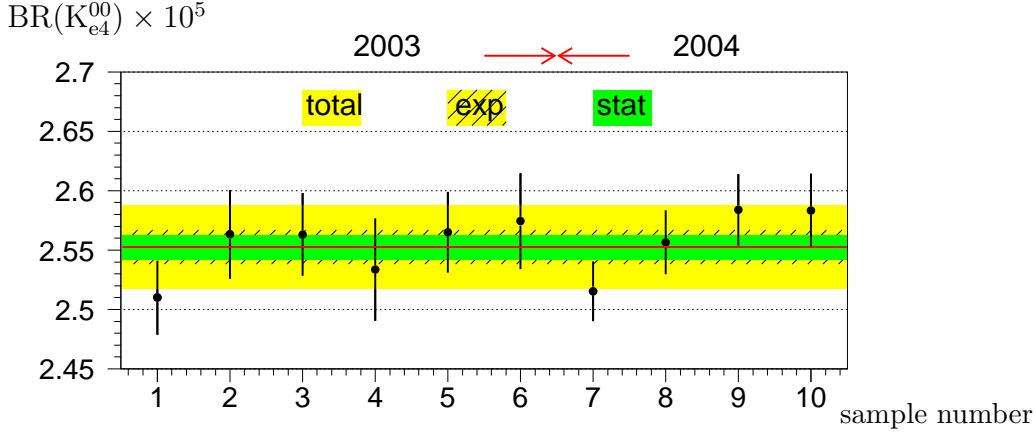}}
\end{picture}
\end{center}
\vspace{-25mm}
\caption{$\KEQO$ branching ratio for ten statistically independent samples 
summed over the two kaon charges. Each error bar corresponds to the 
sample-dependent error of statistical origin (numbers of candidates, 
background, acceptances and trigger efficiencies). The line and the inner band 
correspond to the result of the weighted average and its statistical error. The
hatched band shows the experimental error 
$(\sigma\expe = \sigma\stat \oplus  \sigma\syst)$.
The total error (outer shaded band) includes the external error. The fit 
$\chi^2$ is 6.64 for 9 degrees of freedom (67\% probability) when including all
sample-dependent errors.
\label{fig:brperss}}
\end{figure}

The ratio of partial rates $\Gamma(\KEQO)/ \Gamma(\KTPO)$ is free from the
external error.
The result, including all experimental errors, is obtained as the weighted
average of the ten values obtained from the ten independent subsamples summed 
over both kaon charges:
\begin{eqnarray}\label{eq:ke4rate}
\Gamma(\KEQO) / \Gamma(\KTPO) = (1.449 \pm 0.006\stat \pm 0.006\syst) \times 10^{-3} ,
\end{eqnarray}
which corresponds (using the $\KTPO$ as normalization mode) to the partial rate:
\be
\Gamma(\KEQO) = (2062 \pm 8\stat \pm 8\syst \pm 26\ext) {\rm ~s}^{-1},
\ee
and to the branching ratio:
\be
{\rm BR}(\KEQO) = (2.552 \pm0.010\stat \pm0.010\syst \pm0.032\ext)\times10^{-5},
\label{eq:result}
\ee
where the error is dominated by the external uncertainty from the normalization
mode BR$(\KTPO)= (1.761 \pm 0.022)\%$ \cite{pdg}. The BR$(\KEQO)$ values 
obtained for the ten statistically independent subsamples are shown in 
Fig.~\ref{fig:brperss}, also in agreement with 
the values measured separately for $\KPL$ and $\KMI$:

\begin{displaymath}
{\rm BR}(\KEQP) = ( 2.548 \pm 0.013  )\times10^{-5},~
{\rm BR}(\KEQM) = ( 2.558 \pm 0.018  )\times10^{-5},~
\end{displaymath}
where the quoted uncertainties include statistical and time-dependent
systematic contributions.
The same trigger efficiency values and background to signal ratios as in the 
global analysis have been used to obtain the charge dependent results.

\section{Absolute form factor}
\label{sec:ffabs}
Going back to Eq. (\ref{eq:dg3}) and integrating $d^{3}\Gamma$ over the 
3-dimensional space after substituting $F_{s}$ by its measured parameterization
with $q^2$ and $y^2$ as defined in Eq. (\ref{eq:fxy}), the $\KEQO$
branching ratio, inclusive of radiative decays, is expressed as:
\begin{eqnarray}\label{eq:ffbr}
{\rm BR}(\KEQO) =&
\tau_{K^{\pm}} \cdot (|V_{us}| \cdot  f_{s})^{2} \cdot (1 + \delta_{EM})^{2}
\cdot \int d^{3}\Gamma / (|V_{us}| \cdot  f_{s})^2  ~d \SP ~d \SE ~d\CTE \nonumber \\
=& \tau_{K^{\pm}} \cdot (|V_{us}| \cdot  f_{s} \cdot (1 + \delta_{EM}) )^{2} \cdot I_{3} ,
\end{eqnarray}
where $\tau_{K^{\pm}}$ is the $\KPM$ mean lifetime (in seconds) and
$\delta_{EM}$ a long distance electromagnetic correction to the total rate.
The value of
$f_{s}$ is then obtained from the measured value of BR$(\KEQO)$, 
$\tau_{K^{\pm}}$ and the integration result.
The integral result $I_{3}$ depends on the form factor variation
within the 3-dimensional space (reduced here to a 2-dimensional space as the
$\CTE$ term carries no physics information) and is
computed using the model-independent description as quoted in
Table \ref{tab:fsfe}.
Because of the quadratic dependencies in Eq. (\ref{eq:ffbr}), the relative
uncertainty on $|V_{us}| \cdot  f_{s}$ is only half the relative uncertainty
from the branching ratio, $\tau_{K^{\pm}}$
and phase space integral $I_{3}$.

The statistical and systematic errors of the branching ratio are propagated
while the impact of the limited precision of the form factor description on
the integral $I_3$ is estimated by varying in turn each coefficient $(a,b,c,d)$
by $\pm 1\sigma$. 
External errors affecting the branching ratio and $\tau_{K^{\pm}}$
are propagated to the relative $|V_{us}| \cdot  f_{s}$ uncertainty. The 
additional uncertainty on $|V_{us}|$ is also propagated to the $ f_{s}$ 
measurement (Table \ref{tab:fsys}).

Given the $\KEQO$ branching ratio result from Eq. (\ref{eq:result})
and using the world average 
$\tau_{K^{\pm}} = (1.2380 \pm 0.0021) \times 10^{-8}$ s,
the absolute form factor value is obtained as:
\begin{equation}
(1 + \delta_{EM})\cdot |V_{us}| \cdot  f_{s} = 1.369 \pm 0.003\stat \pm 0.006\syst \pm 0.009\ext\\
\end{equation}
corresponding to
\begin{equation}
 (1 + \delta_{EM}) \cdot f_{s} =  6.079 \pm 0.012\stat \pm 0.027\syst \pm 0.046\ext
\end{equation}
\noindent when using $|V_{us}|=0.2252 \pm 0.0009$~\cite{pdg}. This value shows 
some tension with the corresponding form factor of the $\KEQC$ mode 
$f_{s}^{+-} = 5.705  \pm 0.003\stat \pm 0.017\syst \pm 0.031\ext$ \cite{ke412}.
The observed difference is statistically significant as experimental errors are
mostly uncorrelated. However, a more precise theoretical description of the 
$\KEQO$ mode including radiative, isospin breaking and re-scattering effects 
should be considered before drawing any solid conclusion.

\begin{table}[htp]
\begin{center}
\caption{Summary of the contributions to the $f_{s}$ form factor
uncertainties. 
The external error from $\tau_{K^{\pm}}$ 
may be already accounted for in the
normalization partial rate and should not be counted twice. It has however a
marginal impact on the final error.
\label{tab:fsys}}
\vspace{1mm}
\begin{tabular}{lc}
Source &  $\delta f_s / f_s  (\times 10^{2})$\\
\hline
BR$(\KEQ)$ statistical error & 0.19 \\
BR$(\KEQ)$ systematic error &  0.19 \\
Form factor description (systematic error) &  0.40 \\
Integration method (systematic error) &  0.02\\
\hline
Total experimental error &  0.48   \\
\hline
BR$(\KEQ)$ external error & 0.63 \\
Kaon mean lifetime (external error) & 0.08 \\
$|V_{us}|$  (external error) & 0.40 \\
\hline
Total error (including external errors) & 0.89 \\
\end{tabular}
\vspace{-10mm}
\end{center}
\end{table}

Radiative corrections to $\KEQ$ decays have received only little attention so
far \cite{ke4rad2}, while they have been under study for many years for $\KETC$
($\KTE$) decays which differ from the $\KEQO$ mode by one $\PIo$ in the final
state. Several approaches have been followed within ChPT
\cite{radmk,radseb,radvci}, with Ref. \cite{radvci} quoting
$2\delta_{EM} = (0.10 \pm 0.25)\%$. This could be taken as an indication that 
the $\delta_{EM}$ term is small ($< 1 \times 10^{-3}$) and contributes mainly 
as an additional external relative uncertainty of ${\cal O}(10^{-3})$. A 
dedicated theoretical calculation will be necessary to support this hypothesis 
and could be obtained by adapting a recent evaluation of radiative and isospin 
breaking effects within ChPT in the $\KEQC$ mode \cite{ke4iso}.

\section{Summary}
From a sample of 65210 $\KEQO$ decay candidates with $1\%$ background
contamination, the branching ratio inclusive of 
radiative decays has been measured to be:
\begin{displaymath}
{\rm BR}(\KEQO) =(2.552 \pm0.010\stat \pm0.010\syst \pm0.032\ext)\times10^{-5},
\end{displaymath}
using $\KTPO$ as normalization mode. The $1.4\%$ precision is dominated by the 
external uncertainty from the normalization mode (uncertainties added in
quadrature) and 
represents a factor of 13 improvement over the current world average value,
BR$(\KEQO) = (2.2\pm0.4)\times10^{-5}$. 
The first measurement of the hadronic
form factor has been obtained including its variation in the plane $(\SP ,\SE)$
and providing also evidence for final state charge exchange scattering 
$(\PPI \MPI \ra \PIo \PIo)$ in the $\KEQC$ decay mode below the $2 m_{\pi^+}$
threshold. 
A model independent parameterization has been developed to describe these
variations relative to the form factor value at $\SP =  4 m_{\pi^+}^2, \SE =0.$
Above $\SP = 4 m_{\pi^+}^2$, the relative slope $a$ and curvature $b$
coefficients of a degree-2 series expansion in $q^2 = \SP / 4 m_{\pi^+}^2 -1$
have been obtained together with the relative slope $c$ of a linear dependence 
on $y^2 = \SE / 4 m_{\pi^+}^2$:
\begin{displaymath}
a = \phantom{-}0.149 \pm 0.033\stat \pm 0.014\syst, 
\end{displaymath}
\begin{displaymath}
b =           -0.070 \pm 0.039\stat \pm 0.013\syst, 
\end{displaymath}
\begin{displaymath}
c = \phantom{-}0.113 \pm 0.022\stat \pm 0.007\syst.
\end{displaymath}
These results are in good agreement with those obtained in a high 
statistics measurement of the corresponding form factor of the $\KEQC$ mode.
Below $\SP = 4 m_{\pi^+}^2$, the observed deficit of events is described by a 
cusp-like function
$\sqrt{|q^2/(1+q^2)|}$ with a relative coefficient $d$ and the same linear 
dependence on $y^2$ as above:
\begin{displaymath}
d =           -0.256 \pm 0.049\stat \pm 0.016\syst.
\end{displaymath}
Both total rate and form factor description are used to obtain the absolute
form factor value at $\SP = 4m_{\pi^+}^2 , \SE =0$ ($q^2 = 0, ~y^2 = 0$):
\begin{displaymath}
f_s = 6.079 \pm 0.012\stat \pm 0.027\syst \pm 0.046\ext, 
\end{displaymath}
where the dominating external error comes from uncertainties on the 
normalization mode $\KTPO$ branching ratio, on the mean kaon life time and on
$|V_{us}| = 0.2252 \pm 0.0009$.
An additional external error from a long distance 
electromagnetic correction to the total rate, not available in the literature, 
is expected to contribute at the ${\cal O}(10^{-3})$ relative level.

We are confident that these new and precise measurements will 
prompt fruitful interactions with theorists both in terms of interpretation 
and usage as input to ChPT studies.

\section*{Acknowledgments}
We gratefully acknowledge the CERN SPS accelerator and beam line staff for the
excellent performance of the beam and the technical staff of the participating
institutes for their efforts in maintenance and operation of the detector. We
enjoyed fruitful discussions about $\KEQO$ form factors  
with V.~Bernard and S.~Descotes-Genon in Orsay and M.~Knecht in Marseille. 
\section*{Appendix: Additional information}
Table \ref{tab:detail} gives the definition of the $q^2$ bins used in the 
$\KEQO$ form factor analysis and the input value in each bin.
More information is available upon request to the corresponding author.
\vspace{-1mm}
\begin{table}[htp]
\begin{center}
\begin{tabular}{l|rrc}
 bin    & \multicolumn{1}{c}{$\MPP$ range} & \multicolumn{1}{c}{$q^2$} & $(N_s - N_{b}(s))/N_{MC}$ \\
 number & \multicolumn{1}{c}{$(\MEVcc)$}  & barycenter &                     \\
\hline
1 & $2m_{\pi^0}- 275.57$ & $-0.0413$ & $0.9106 \pm 0.0171$ \\
2 & $275.57 - 279.14$    & $-0.0123$ & $0.9353 \pm 0.0175$ \\
3 & $279.14 - 285.09$    & $ 0.0217$ & $0.9944 \pm 0.0133$ \\
4 & $285.09 - 290.78$    & $ 0.0641$ & $1.0213 \pm 0.0136$ \\
5 & $290.78 - 296.86$    & $ 0.1077$ & $1.0065 \pm 0.0133$ \\
6 & $296.86 - 303.01$    & $ 0.1541$ & $1.0417 \pm 0.0139$ \\
7 & $303.01 - 309.56$    & $ 0.2032$ & $1.0838 \pm 0.0145$ \\
8 & $309.56 - 317.32$    & $ 0.2598$ & $1.0511 \pm 0.0140$ \\
9 & $317.32 - 326.48$    & $ 0.3284$ & $1.0705 \pm 0.0142$ \\
10 & $326.48 - 338.36$   & $ 0.4159$ & $1.0531 \pm 0.0141$ \\
11 & $338.36 - 355.53$   & $ 0.5383$ & $1.0909 \pm 0.0146$ \\
12 & $      > 355.53$    & $ 0.8004$ & $1.1293 \pm 0.0148$ \\
\end{tabular}
\end{center}
\vspace{-15mm}
\caption{Description of the 12 bins of unequal width in $q^2$:
bin range in $\MPP$, corresponding $q^2$ barycenter position, ratio of numbers 
of events from data (background subtracted) and simulation (constant form 
factor). The errors are statistical only. The boundary between bins 2 and 3 
corresponds to $\MPP = 2m_{\pi^+}$ ($q^2 = 0$).
\label{tab:detail} }
\end{table}


\end{document}